\documentclass[5p,sort&compress,longtitle]{elsarticle}

\usepackage{hyperref}
\usepackage{caption}
\usepackage{subcaption}
\usepackage{amsmath}
\usepackage{amssymb}
\usepackage{fixltx2e}
\usepackage{tabularx}
\usepackage{graphicx}
\usepackage{units}
\usepackage{color}

\renewcommand{\eqref}[1]{(Eq.~\ref{#1})}

\begin{document}

\hyphenation{brems-strahl-ung}

\begin{frontmatter}
\title{Energy Reconstruction Methods in the IceCube Neutrino Telescope}
\journal{Journal of Instrumentation}
\author[Adelaide]{M.~G.~Aartsen}
\author[MadisonPAC]{R.~Abbasi}
\author[Zeuthen]{M.~Ackermann}
\author[Christchurch]{J.~Adams}
\author[Geneva]{J.~A.~Aguilar}
\author[MadisonPAC]{M.~Ahlers}
\author[Erlangen]{D.~Altmann}
\author[MadisonPAC]{C.~Arguelles}
\author[MadisonPAC]{J.~Auffenberg}
\author[Bartol]{X.~Bai\fnref{SouthDakota}}
% ...
\address[Adelaide]{School of Chemistry \& Physics, University of Adelaide, Adelaide SA, 5005 Australia}
% ...
\address[Christchurch]{Dept.~of Physics and Astronomy, University of Canterbury, Private Bag 4800, Christchurch, New Zealand}
% ...
\address[Erlangen]{Erlangen Centre for Astroparticle Physics, Friedrich-Alexander-Universit\"at Erlangen-N\"urnberg, D-91058 Erlangen, Germany}
\address[Geneva]{D\'epartement de physique nucl\'eaire et corpusculaire, Universit\'e de Gen\`eve, CH-1211 Gen\`eve, Switzerland}
% ...
\address[MadisonPAC]{Dept.~of Physics and Wisconsin IceCube Particle Astrophysics Center, University of Wisconsin, Madison, WI 53706, USA}
% ...
\address[Bartol]{Bartol Research Institute and Dept.~of Physics and Astronomy, University of Delaware, Newark, DE 19716, USA}
% ...
\address[Zeuthen]{DESY, D-15735 Zeuthen, Germany}
\cortext[cor]{Corresponding author}
\fntext[SouthDakota]{Physics Department, South Dakota School of Mines and Technology, Rapid City, SD 57701, USA}
\fntext[Bari]{also Sezione INFN, Dipartimento di Fisica, I-70126, Bari, Italy}
\fntext[Goddard]{NASA Goddard Space Flight Center, Greenbelt, MD 20771, USA}

\begin{abstract}
Accurate measurement of neutrino energies is essential to many of the scientific goals of large-volume neutrino telescopes. The fundamental observable in such detectors is the Cherenkov light produced by the transit through a medium of charged particles created in neutrino interactions. The amount of light emitted is proportional to the deposited energy, which is approximately equal to the neutrino energy for $\nu_e$ and $\nu_\mu$ charged-current interactions and can be used to set a lower bound on neutrino energies and to measure neutrino spectra statistically in other channels. Here we describe methods and performance of reconstructing charged-particle energies and topologies from the observed Cherenkov light yield, including techniques to measure the energies of uncontained muon tracks, achieving average uncertainties in electromagnetic-equivalent deposited energy of $\sim 15\%$ above 10 TeV.
\end{abstract}


\end{frontmatter}

\section{Introduction}
\label{sec:energyreco}

The IceCube neutrino observatory \cite{daqpaper,2006APh....26..155I} is a cubic kilometer photomultiplier (PMT) array embedded in glacial ice at the geographic South Pole. The complete array is made of 5160 downward-facing Hamamatsu R7081 photomultipliers deployed on 86 vertical strings at depths between 1450 and 2450 meters in the icecap. IceCube detects neutrinos by observing Cherenkov light induced by charged particles created in neutrino interactions as they transit the ice sheet within the detector; the energy and momentum of these charged particles reflect the energy and momentum of the original neutrino. 

At the TeV energies typical of such neutrino telescopes, the primary neutrino interaction channel is deep-inelastic scattering with nuclei in the detector material. In both neutral and charged-current interactions, a shower of hadrons is created at the neutrino interaction vertex. In charged-current interactions, this shower is accompanied by an outgoing charged lepton. This lepton, in particular for electrons, may also lose energy rapidly and itself trigger another overlaid shower. Cherenkov light is radiated by this primary lepton and any accompanying showers with a total amplitude proportional to the integrated path length of charged particles above the Cherenkov threshold. This, in turn, is proportional to the total energy of these particles \cite{cascade_light}.

The light production from electromagnetic (EM) showers is both maximal and has low variance with respect to deposited energy \cite{cascade_light}. As such, it forms a natural unit of reconstructed shower energy. This electromagnetic-equivalent energy, in conjunction with the energies of detected outgoing leptons, can then be used to infer the energy of the original neutrino. In identifiable charged-current interactions (e.g. $\nu_\mu$ CC events), neutrino energy resolution for events where the interaction vertex is observed is in principle limited only by detector resolution. For neutral-current events, neutrino energy spectra can be inferred statistically. A similar method can be used to estimate energy spectra for events where not all charged particles are contained within the detector \cite{photorec_icrc}, such as muons produced in charged-current $\nu_\mu$ interactions an unknown distance outside the detector.

\begin{figure}
\begin{centering}
\includegraphics[width=\linewidth]{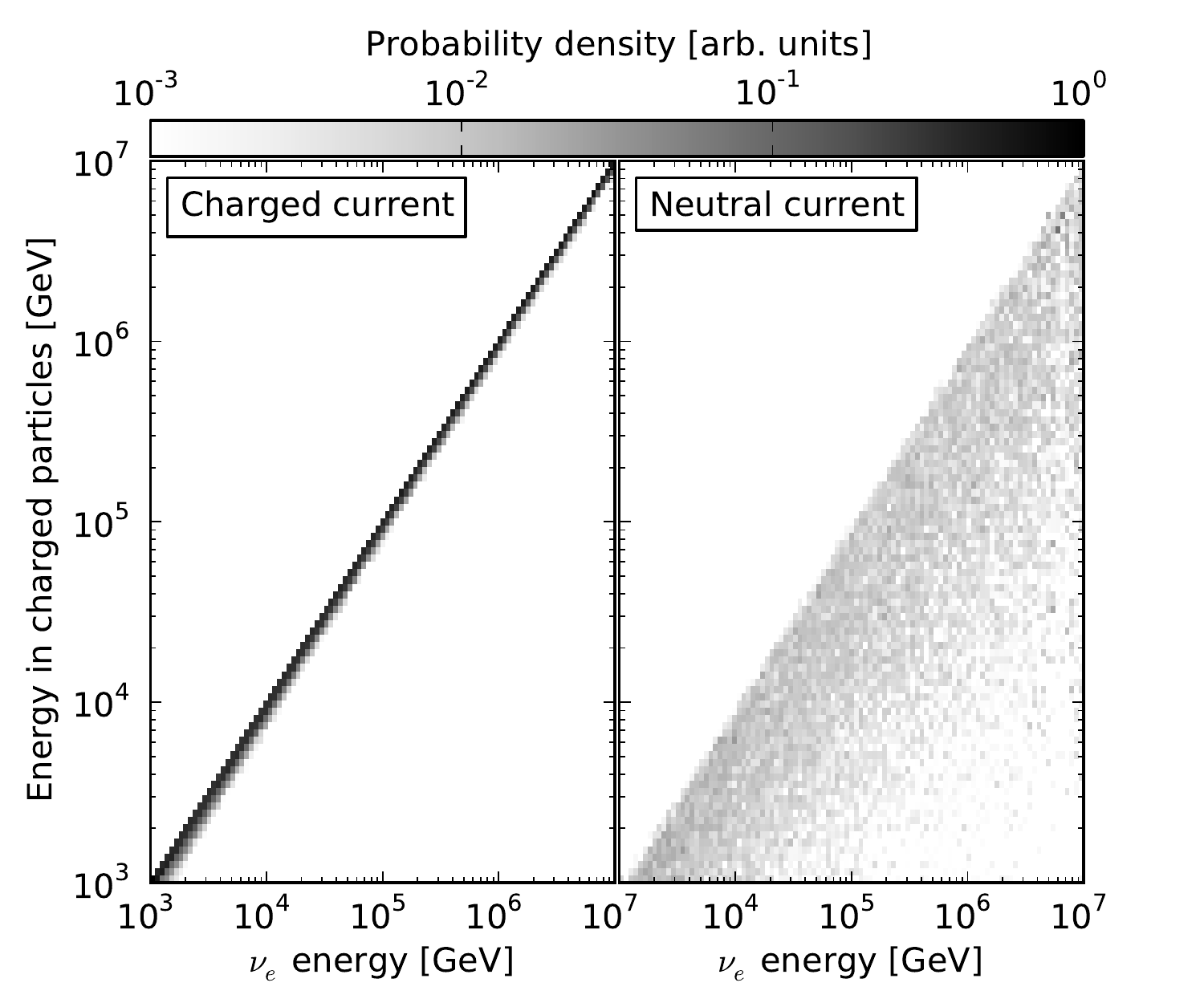}
\caption[Deposited energy of $\nu_e$]{
Energy deposited in Cherenkov-radiating particles by deep-inelastic $\nu_e$-nucleon scatterings in ice \cite{gandhi_xsections}. In charged-current interactions within the detector this deposited energy is very nearly equal to the neutrino energy. The larger spread to lower deposited energies in the right panel is due to neutral-current scattering; the rate of neutral-current interactions is approximately 3 times smaller than that of charged-current interactions in the energy region of interest \cite{css_crosssections}. 
}
\label{fig:nue_visible_energy}
\end{centering}
\end{figure}

Electromagnetic showers are produced in $\nu_e$ charged-current interactions from the outgoing electron, in $\tau$ decays, and along muon tracks from muon bremsstrahlung and pair production interactions. At high energies ($\gtrsim 1$ TeV) such stochastic showers dominate light output from muons. Electromagnetic showers have nearly identical light deposition patterns independent of energy \cite{cascade_light} from IceCube's threshold until the onset of the LPM effect \cite{Landau:1953gr,Migdal:1956tc,2010PhRvD..82g4017G} at energies of many PeV. Hadronic showers, produced as the sole signature of neutral-current neutrino scatterings, as a component of charged-current scatterings, and from muon photonuclear interactions, provide very similar profiles but with a suppressed light yield. These also have larger statistical variance in the relationship between energy and Cherenkov light due to the presence of more neutral particles \cite{wiebusch_thesis}. Both the relative suppression and variance decrease with energy as more neutral pions directly feed the electromagnetic part of the shower \cite{Gabriel1994336}; a 100 \unit{GeV} (100 \unit{TeV}) hadronic cascade produces on average 74\% (89\%) as much Cherenkov light as a purely electromagnetic cascade of the same energy, with shower-to-shower variations of 17\% (6\%) relative to the average \cite{wiebusch_thesis}.

\begin{table*}
\begin{tabular}{c|c|c|c|c}
Interaction & Signature & $E_{vis}/E_\nu$; $E_\nu = 1$ TeV & $E_\nu = 10$ TeV & $E_\nu = 100$ TeV \\
\hline
$\nu_e + N \rightarrow e + had.$ & Cascade & 94\% & 95\% & 97\% \\
$\nu_\mu + N \rightarrow \mu + had.$ & Track (+ Cascade) & 94\% & 95\% & 97\% \\
$\nu_\tau + N \rightarrow \tau + had. \rightarrow had.$ & Cascade/Double Bang & $< 94\%$ & $< 95\%$ & $< 97\%$ \\
$\nu_\tau + N \rightarrow \tau + had. \rightarrow \mu + had.$ & Cascade + Track & $< 94\%$ & $< 95\%$ & $< 97\%$ \\
\hline
$\nu_l + N \rightarrow \nu_l + had.$ & Cascade & 33\% & 30\% & 23\% \\
\end{tabular}
\caption{
	Neutrino interactions with nucleons in IceCube. $E_{vis}$ denotes the median fraction of the neutrino energy deposited in any present primary lepton and in the EM-equivalent energy of a hadronic cascade at the vertex. In charged-current interactions (top section of table), nearly all the energy of the interacting neutrino is deposited in such light-producing particles. In neutral-current interactions (bottom), a large fraction of the neutrino energy leaves with the outgoing neutrino (Fig. \ref{fig:nue_visible_energy}) \cite{gandhi_xsections}. Note that some of $E_{vis}$ may escape the detector: muon tracks at these energies have lengths of multiple kilometers, and $\tau$ leptons will decay before ranging out, depositing only a fraction of $E_{vis}$ in the detector. Events in IceCube are observed as a combination of cascades (near-pointlike particle showers) and long tracks, as are left predominantly by muons. ``Double Bang'' refers to two cascades joined by a short track, a signature of charged-current $\nu_\tau$ interactions at high energies ($\gtrsim 1\,\,\unit{PeV}$) where the separate production and decay cascades of the $\tau$ are resolvable in IceCube. Due to the long lengths of muon tracks above 1 \unit{TeV}, most observed neutrino-induced muons have production vertices outside the detector and the initial hadronic cascade is not observed.}
\label{table:signatures}
\end{table*}

\begin{figure*}
\begin{centering}
\begin{subfigure}{0.48\linewidth}
	\includegraphics[width=\linewidth]{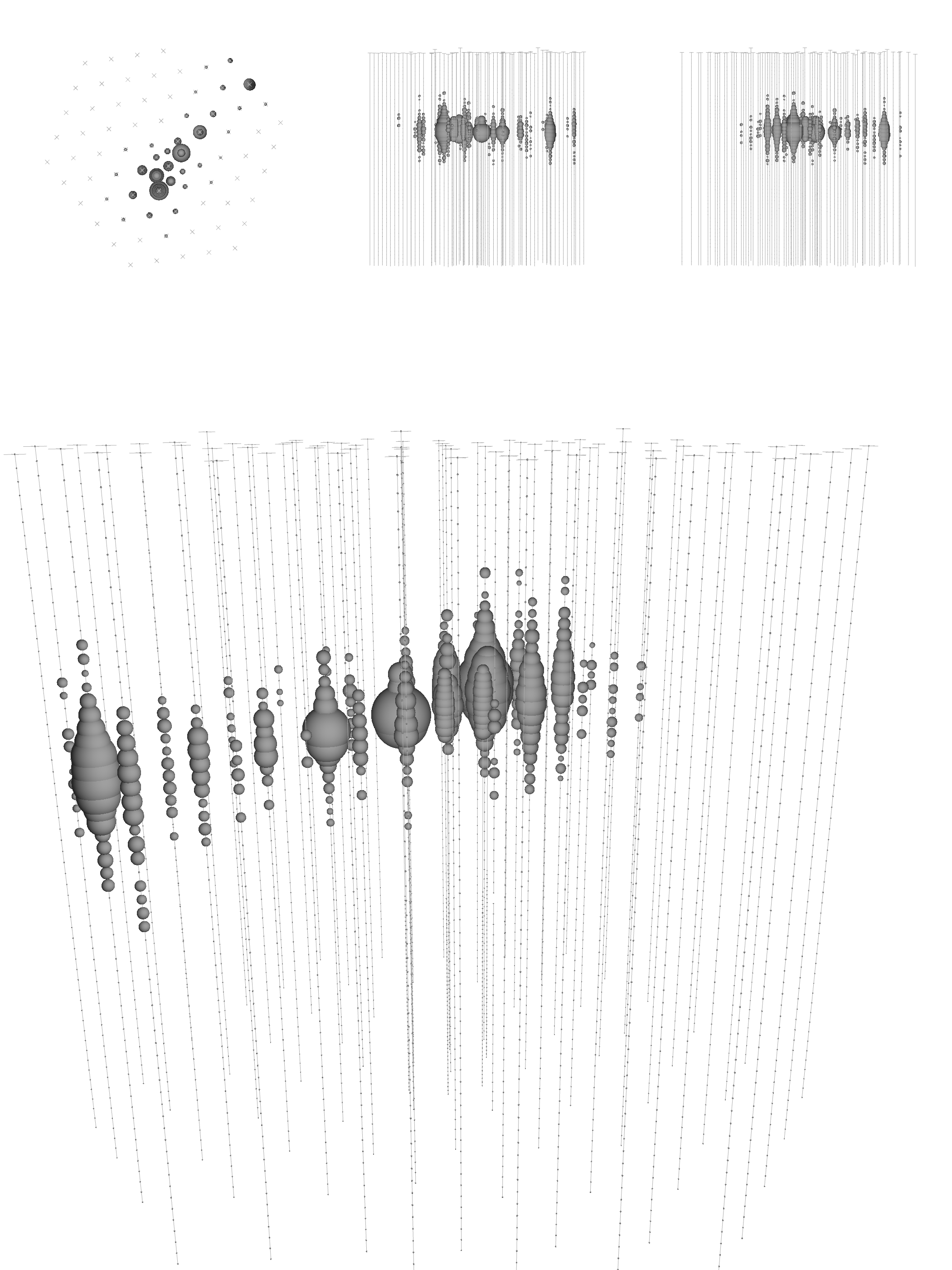}
	\caption{A muon that started in the detector and deposited 74 TeV before escaping, carrying away its remaining energy.}
\end{subfigure}
\begin{subfigure}{0.48\linewidth}
	\includegraphics[width=\linewidth]{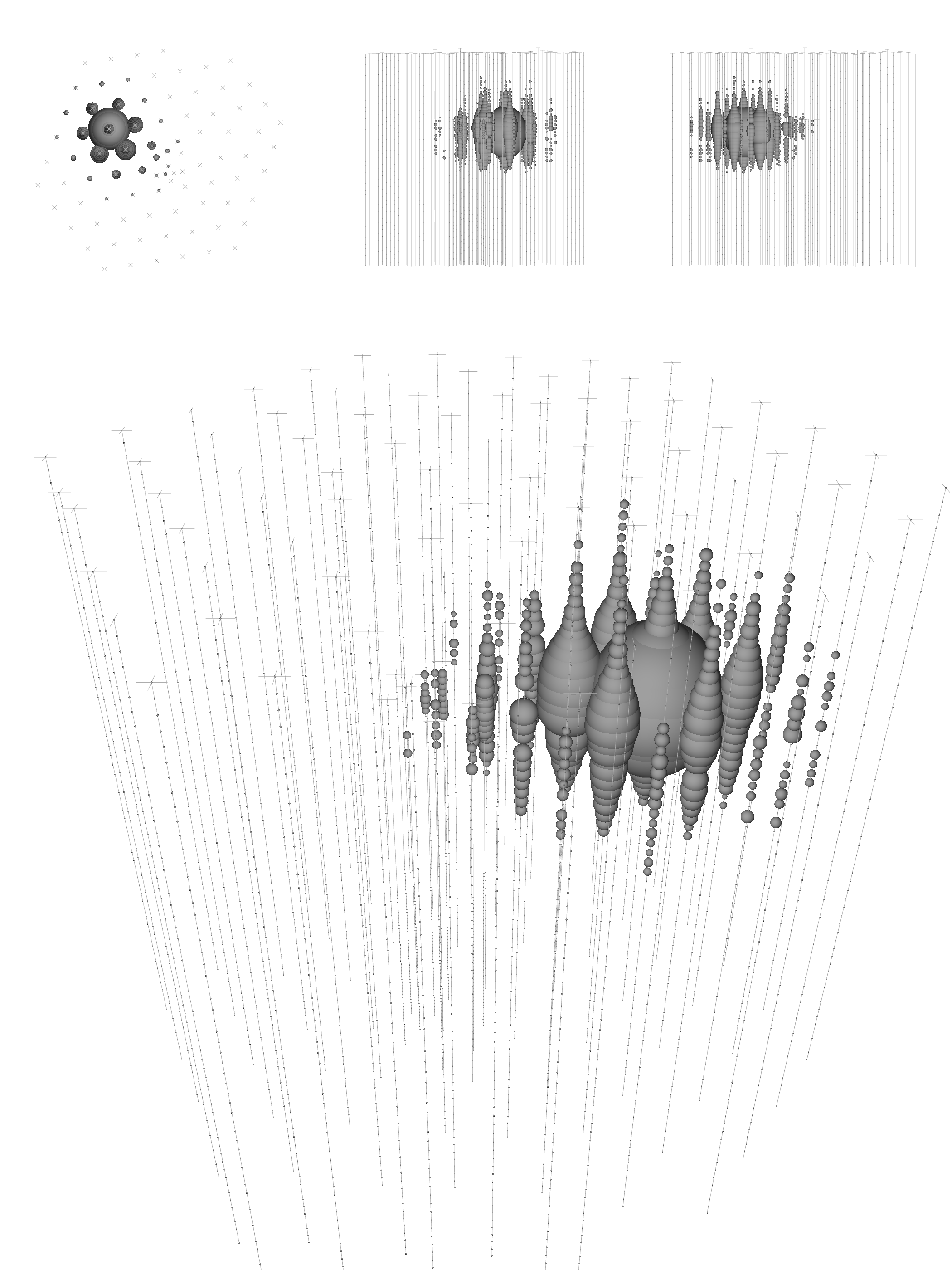}
	\caption{A cascade that deposited 1070 TeV in the detector. Its energy can be determined directly since the cascade is completely contained in the instrumented volume.}
\end{subfigure}
\caption[Examples of event topologies]{Examples of neutrino event topologies in IceCube from \cite{hese_paper}. Each panel is a schematic view of the detector, with each photomultiplier represented by a sphere whose volume is proportional to the collected charge. The smaller upper panels show projections of the detector along its $z$, $x$, and $y$ axes, respectively.}
\label{fig:event_topologies}
\end{centering}
\end{figure*}

As a result of the dominance of the electromagnetic shower, the energy deposited in the detector is nearly identical to the neutrino energy for charged-current $\nu_e$ interactions (Table~\ref{table:signatures}, Fig.~\ref{fig:nue_visible_energy}), but the distribution is quite broad for neutral-current interactions, in which the outgoing neutrino carries a large and highly variable fraction of the energy out of the detector.

\begin{figure}
\begin{centering}
\includegraphics[width=\linewidth]{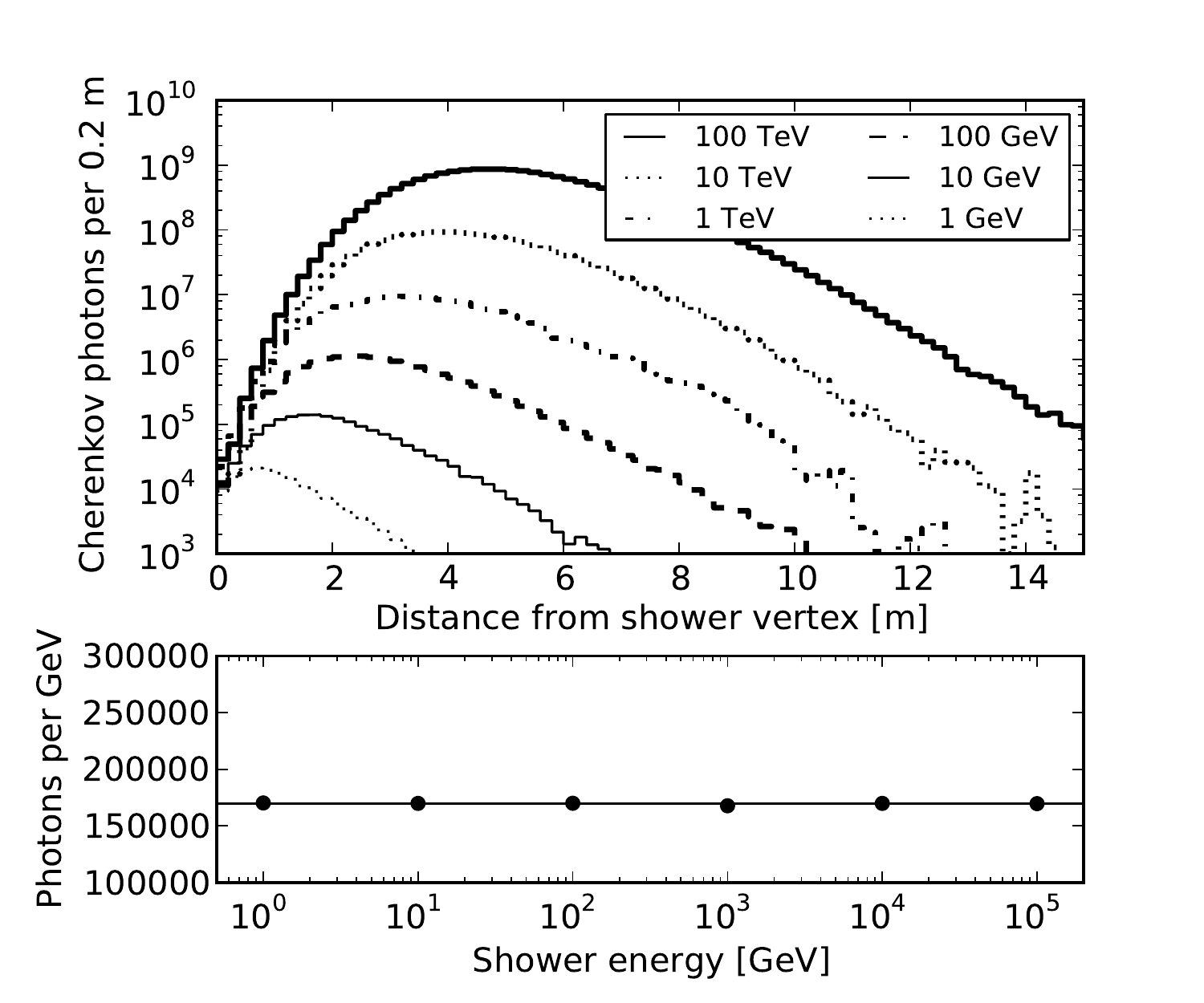}
\caption[Longitudinal profiles of electromagnetic cascades]{
Longitudinal distributions of electromagnetic cascades in ice simulated with GEANT4 \cite{GEANT4}. The total light output of the cascade is directly proportional to its energy, while the length of the shower increases only logarithmically with energy. The angular profile of light emission is nearly identical across the entire energy range \cite{cascade_light}. The range in positions of the shower maximum (between 1 and 5~meters from the vertex here) is small enough that the showers can be considered pointlike on the scale of IceCube instrumentation. Hadronic showers have nearly identical properties when viewed with IceCube \cite{wiebusch_thesis}.
}

% NB: the bin counts are the number of Cherenkov photons between 300 and 600 nm

\label{fig:em_longitudinal_profile}
\end{centering}
\end{figure}

IceCube events have two basic topologies: tracks and cascades (Table \ref{table:signatures}, Fig. \ref{fig:event_topologies}). Tracks are made predominantly by muons, either from cosmic-ray air showers or $\nu_\mu$ charged-current interactions. Cascades are those events without visible muon tracks and are formed by particle showers near the neutrino vertex. These are produced in $\nu_e$ charged-current and all-flavor neutral-current interactions. The particle showers in cascade events have typical lengths of 10 m (Fig.~\ref{fig:em_longitudinal_profile}), which are not in general resolvable with the 17 m vertical inter-PMT spacing and 125 meter horizontal inter-string spacing of the IceCube array. As a result, it is not possible to separate $\nu_e$ charged-current interactions from neutral-current interactions.

For both track and cascade events, we discuss the general approach to energy reconstruction and provide examples of the most commonly used algorithms. All energy reconstruction methods described here are based on the linearity of light yield with energy loss and use the common likelihood model described below. Performance data provided are meant to characterize the behavior of the energy reconstruction only and, except when noted otherwise, are given assuming that the topology of the events, in particular direction and position, are known exactly. This controls for uncertainties induced by positional and directional reconstructions and shows the intrinsic uncertainties of the reconstruction being discussed. Although the resolutions shown here are typical of energy resolutions in IceCube physics analyses, some variation should be expected due to uncertainties from topology reconstructions, which will tend to worsen the resolution, as well as from selection of well-reconstructed events in the analyses, which tends to improve it. To show typical performance in physics analyses, we include the final-level resolutions of the algorithm being discussed for a recent example IceCube analysis at the end of each section.

In addition to performance of energy reconstruction in simulation, we also discuss relevant calibration issues in data. Accurate measurement of energies requires correct inference of incident Cherenkov photon fluxes from the digitized photomultiplier signals. We demonstrate this here using verification of the PMT anode current reconstruction and single-photoelectron calibration (Sec.~\ref{sec:photon_counting}), the PMT quantum efficiency, optical transmissivity of the DOM, and overall energy scale using low-energy muons (Sec.~\ref{sec:energyscale}), and linearity of response over a wide range of photon fluxes using a dedicated calibration laser (Sec.~\ref{sec:standardcandle}). These establish the validity of the models used in the reconstructions and complement previous calibration measurements of the IceCube instrumentation \cite{daqpaper,2010NIMPA.618..139A}.

\section{Likelihood Model}

The near-constant light emission profile of both electromagnetic and hadronic showers and the linear scaling of light output with energy allow the use of such showers as fundamental units of energy reconstruction by scaling the expected light output of a simulated event (a ``template'') to match observed data. We then estimate a shower's energy deposition $E$ by comparing the observed number of photons in a PMT $k$ to the expectation $\Lambda$ for a template event with some reference energy (usually 1~GeV) \cite{photorec_icrc}. The template functions are typically evaluated from tabulated \cite{photospline} Monte Carlo simulation \cite{Lundberg:2007p2} of light propagation in the ice sheet \cite{aha,spice}, although limited-accuracy analytic approximations (Sec.~\ref{sec:analytic_approximation}) or direct real-time Monte Carlo simulation can also be used. These template functions take into account the expected detector response as well as position-dependent light propagation properties due to wind-deposited particulate layers deep in the glacier \cite{spice} and will be described in more detail in the following section.

The number of detected photons is expected to follow a Poisson distribution with mean $\lambda=\Lambda E$.  Then the likelihood $\mathcal{L}$ for an energy $E$ resulting in $k$ detected photons from an event producing $\Lambda$ photons per unit energy can be evaluated as follows:

\begin{equation}
\label{eq:poissonllh}
\begin{array}{rl}
\mathcal{L}	& = \frac{\lambda^k}{k!} \cdot e^{- \lambda} \\
		& \lambda \rightarrow E \Lambda \\
		& = \frac{\left ( E \Lambda \right )^k}{k!} \cdot e^{- E \Lambda} \\
\ln \mathcal{L}	& = k \ln \left (E \Lambda \right ) - E \Lambda - \ln \left (k! \right ).
\end{array}
\end{equation}
Maximizing this with respect to energy, and adding the contributions from all DOMs (digital optical modules):

\begin{equation}
\label{proof:energyreco}
\begin{array}{rl}
0 = \frac{\partial \sum \ln \mathcal{L}}{\partial E} & = \sum_{\textrm{DOMs}\,\,j} \left ( k_j \Lambda_j / E \Lambda_j - \Lambda_j \right )\\
		& = \sum k_j/E - \sum \Lambda_j \\
\therefore E	&= \sum k_j / \sum \Lambda_j.
\end{array}
\end{equation}

The generalization allowing additional contributions (e.g. PMT noise) is to replace the substitution $\lambda = E \Lambda$ in Eq.~\ref{eq:poissonllh} by $\lambda = E \Lambda + \rho$, where $\rho$ is the expected number of noise photons. The likelihood \eqref{eq:poissonllh} then becomes:

\begin{equation}
\label{eq:energyrecownoise}
\ln \mathcal{L}	= k \ln \left (E \Lambda + \rho \right ) - \left (E \Lambda + \rho \right ) - \ln \left (k! \right ).
\end{equation}
Maximizing with respect to $E$, as in Eq. \ref{proof:energyreco}:

\begin{equation}
\begin{array}{rl}
0		& = \sum \left ( k_j \Lambda_j / \left (E \Lambda_j + \rho_j \right ) - \Lambda_j \right) \\
\sum \Lambda_j	& = \sum k_j \Lambda_j / \left (E \Lambda_j + \rho_j \right ).
\end{array}
\end{equation}
Unlike Eq. \ref{proof:energyreco}, this does not have a closed form solution for $E$ since $\Lambda$ no longer cancels in the first term and $E$ can therefore not be factored out. Solutions can, however, be easily obtained using gradient-descent numerical minimization algorithms.

Timing can also be included in this formulation by dividing the photon time arrival distributions (see Sec.~\ref{sec:photon_counting}) for bright events into multiple time bins and interpreting $k$ and $\Lambda$ as the light per time bin instead of per PMT. The formulation of the Eq.~\ref{eq:energyrecownoise} is identical under this change. Use of detailed timing provides small increases in energy reconstruction performance in the single-source case. It is primarily only important in more complex situations such as using unfolding to estimate energies of multiple simultaneously emitting light sources (Sec.~\ref{sec:millipede}). Timing is also used when only parts of the observed charge distribution are usable due to, for example, saturation in the photomultipliers and digitizers since it allows the saturated portions of the readout to be masked out during the fit.

\section{Methods to Compute Light Yields}

The ability to reconstruct $E$ relies on correct computation of the light-yield scaling function $\Lambda$. This function depends on the positions of the observing photomultiplier ($\vec x_p$), the position of the event vertex ($\vec x_\nu$), the orientation of the event $(\theta, \phi)$, and, when using timing information, the time the particle was at $\vec x_\nu$ and the time of observation. The typical observation distance a few scattering lengths away from the source, the complex wavelength dependence of light propagation, and the inhomogeneous optical properties of the ice \cite{spice,aha} make a precise analytic form for $\Lambda$ impossible. For applications requiring speed more than accuracy, an approximate form for the light yield can be derived. Final reconstructions depend on tabulated results of Monte Carlo simulation of in-ice light propagation \cite{Lundberg:2007p2} smoothed with a multi-dimensional spline surface \cite{photospline}. Direct use of Monte Carlo, without pre-tabulation, is also possible but computationally prohibitive in almost all applications.

\subsection{Analytic Approximation of Expected Light Yields}
\label{sec:analytic_approximation}

An approximation for point-like spherically-symmetric emission (an approximation to an electromagnetic shower) or uniform emission at the Cherenkov angle along an infinite track (an approximation to a minimum-ionizing muon) is possible when computational speed is essential. In the limit of little scattering, photons propagate in straight lines away from the source and photon density decreases exponentially with distance due to absorption. Near the source this results in a $1/r^2$ dependence for a point source and $1/r$ for a cylindrical source like a muon ($r$ being the distance of closest approach). At larger distances, propagation of photons enters a diffusive regime in which the motion of the photons can be approximated with a random walk, changing the photon density to $\exp{(-r/\lambda_p)}/r$ for a point source and $\exp{(-r/\lambda_p)}/\sqrt{r}$ for a cylindrical source. The characteristic ``propagation'' length ($\lambda_p$) is defined via the absorption and effective scattering lengths, $\lambda_a$ and $\lambda_e$: $\lambda_p=\sqrt{\lambda_a\lambda_e/3}$ \cite{aha}. The two descriptions (near and far) can be combined in a single expression by an empirical functional form matching both limiting cases:

\begin{equation}
\begin{array}{rl}
\mu(r)=&n_0A\cdot{1\over 4\pi}e^{-r/\lambda_p}{1\over\lambda_cr\tanh(r/\lambda_c)}, \\
 & {\sf where} \quad \lambda_c={\lambda_e\over 3\zeta}, \quad \zeta=e^{-\lambda_e/\lambda_a} \quad \mbox{(point source),}\\
\\
\mu(r)=&l_0A\cdot{1\over 2\pi\sin\theta_c} e^{-r/\lambda_p}{1\over \sqrt{\lambda_\mu r} \tanh\sqrt{r/\lambda_\mu}}, \\
& {\sf where} \quad \sqrt{\lambda_\mu}={\lambda_c\over\sin\theta_c} \sqrt{2\over\pi\lambda_p} \quad \mbox{(track).}
\end{array}
\label{eq:dimaparam}
\end{equation}

\begin{figure}
\begin{center}
	\includegraphics[width=\linewidth]{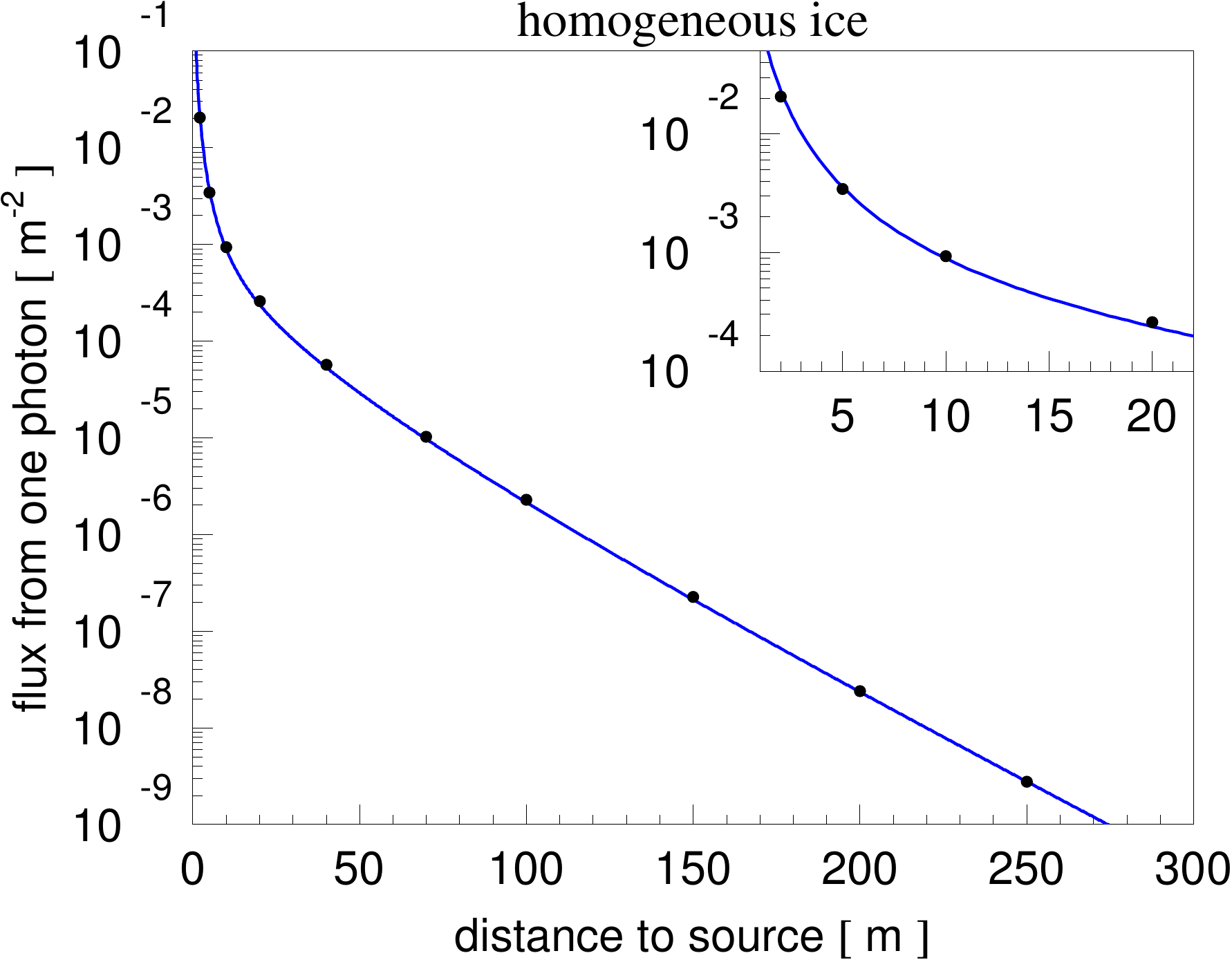}
	\caption{Approximations to the observed light level as a function of distance for point light sources. Lines show Eq.~\ref{eq:dimaparam}. Results from a Monte Carlo simulation of photon tracking are shown with black points. The insert at the top right shows small distances. These plots are made assuming homogeneous optical properties of the ice, whereas the actual glacial ice has depth-varying properties \cite{spice}.}
	\label{fig:PARvsPPC}
\end{center}
\end{figure}

In these expressions, $n_0$ is the number of photons emitted by a point source and $l_0$ the number of photons per meter from the uniform track source. The quantity $A$ is the effective photon collection area of the receiving sensor and $\theta_c$ is the Cherenkov angle. We have verified these formulae with Monte Carlo photon tracking \cite{Chirkin:2013tma}. For the typical values of $\lambda_a$=98 m, $\lambda_e$=24 m, the description can be further improved by using a fitted value of $\lambda_p^*=\lambda_p/1.07$, and a corresponding increase in normalization by 26\% (Fig. \ref{fig:PARvsPPC}).
Since the optical properties of the ice vary with depth, the values of $1/\lambda_a$, $1/\lambda_e$ are taken as averages of the local values between the emitter and receiver.

In practice, considerable errors are introduced by the analytic expressions above due to various effects that they ignore: for example, the directionality of the cascade, which is not really an isotropic source of photons. These approximations result in poor performance when using Eq.~\ref{eq:poissonllh}, which assumes a high-quality representation of the expected light output and the dominance of statistical uncertainties. When using these analytic approximations (e.g. for reasons of computational speed), a variant on Eq.~\ref{eq:poissonllh} is required with wider tails to cover the approximation uncertainty. We incorporate this by convolving Eq.~\ref{eq:poissonllh} with a probability distribution $G$ on the mean light-yield $\lambda$, which was chosen empirically to be:

\begin{equation}
G_{\mu}(x)={\mbox{const.}\over x}\cdot\left(e^{-wy}+{(y/\sigma)^2}\right)^{-1}, \, \mbox{with } y=\ln{x\over\mu}.
\end{equation}
The parameter $w$ is a ``skewness'' parameter, which allows for larger over-fluctuations (e.g. in case of a large bremsstrahlung loss along a muon track---see Sec.~\ref{sec:muons}).

\subsection{Spline Tables}
\label{sub:spline_tables}

\begin{figure*}
\begin{centering}
\begin{subfigure}[t]{0.48\linewidth}
	\includegraphics[width=\linewidth]{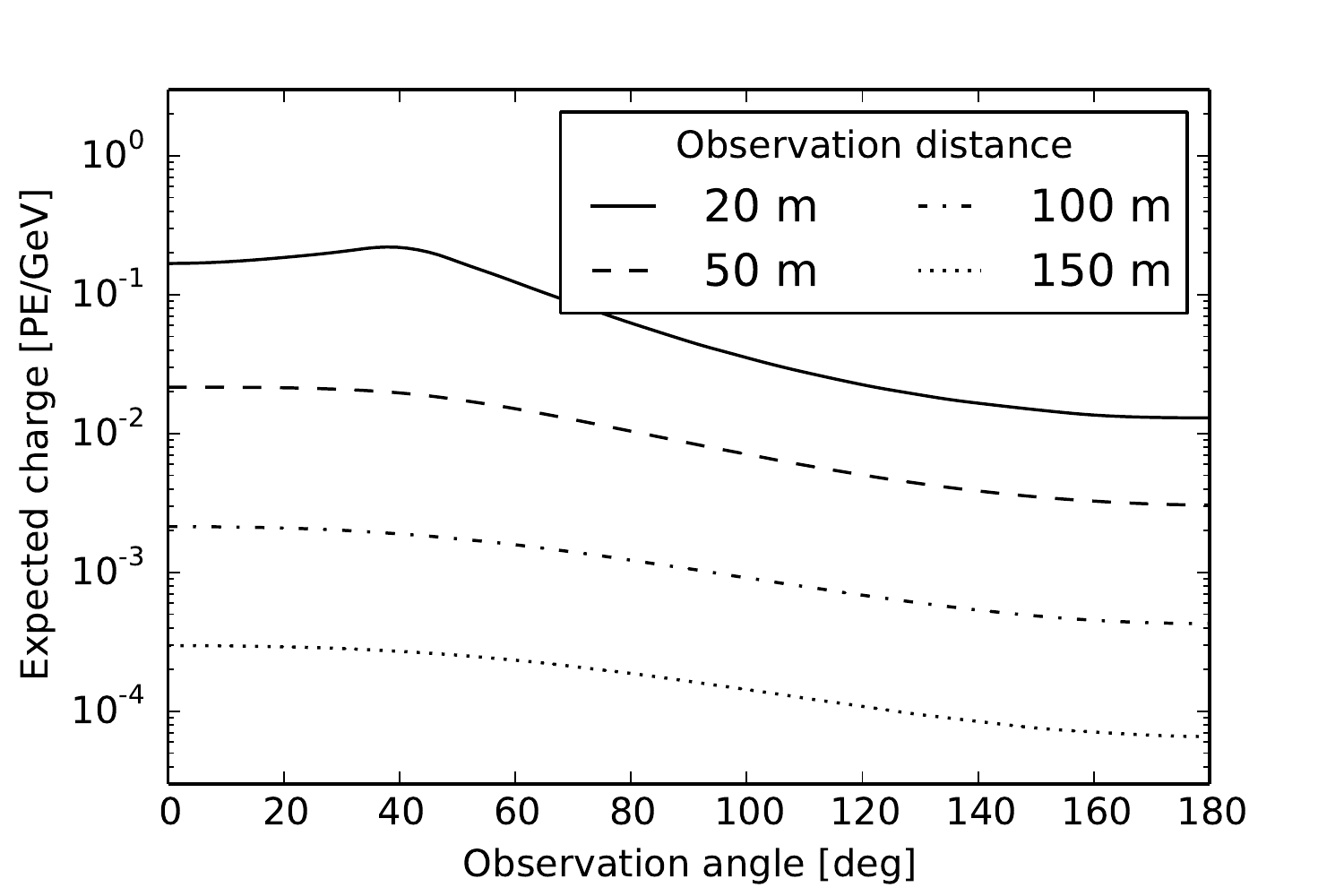}
	\caption{Total observed light level as a function of radial distance from the source and observation angle with respect to the source direction. While scattering in the ice washes out the peak at the Cherenkov angle, the direction of the source remains visible as an asymmetry even at large distances.}
\end{subfigure}
\begin{subfigure}[t]{0.48\linewidth}
	\includegraphics[width=\linewidth]{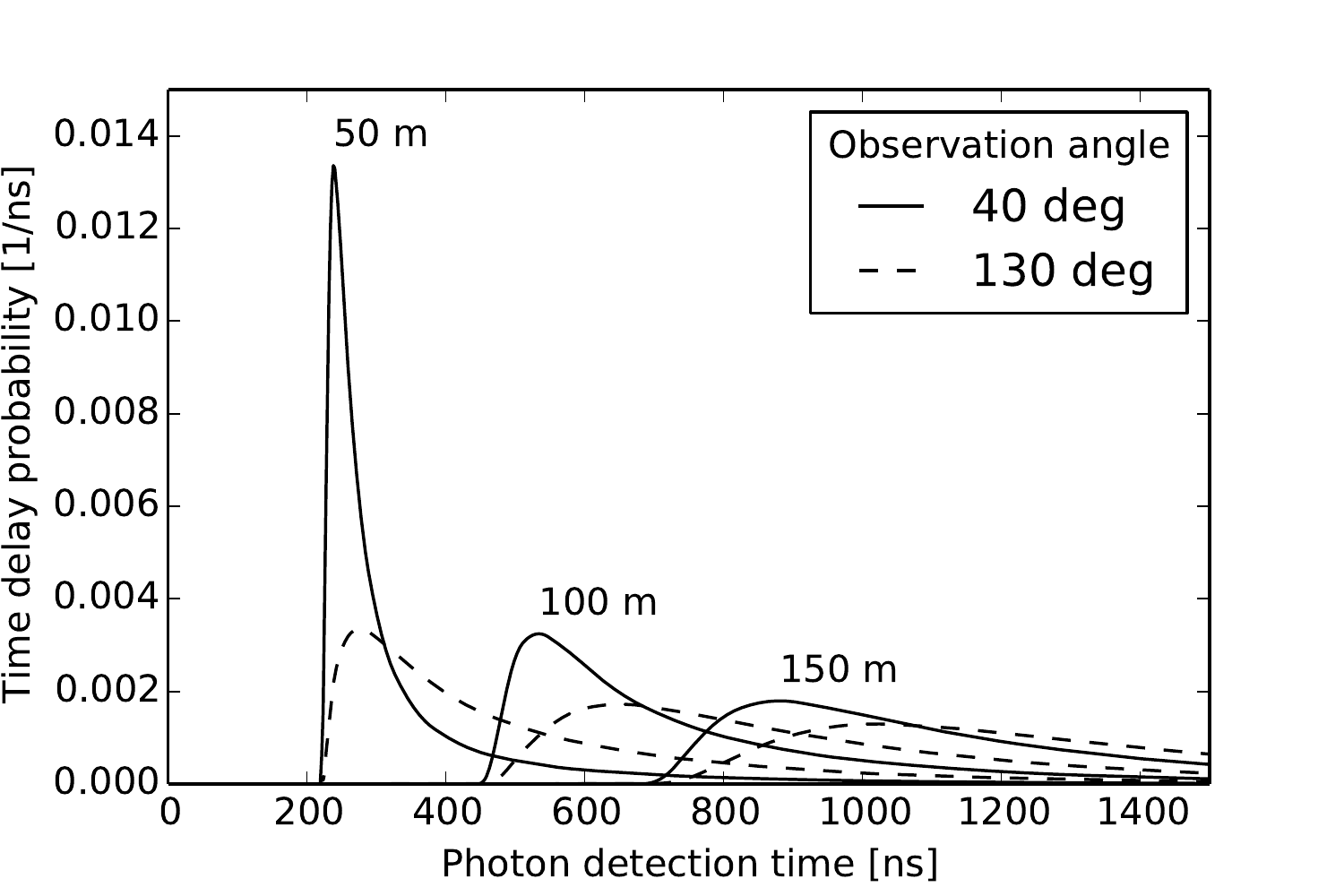}
	\caption{Normalized time distribution of detected photons at different distances for two observation angles. Photons detected at the Cherenkov angle have generally experienced the least scattering, and so are detected earlier and more closely bunched in time than those detected at other angles.}
\end{subfigure}
\caption[]{Distribution of detectable photons obtained from a Monte Carlo simulation of a horizontal, 1 GeV electromagnetic cascade in the upper part of the IceCube detector. Both the number and time distribution of photons depend on the direction of the cascade, here oriented in the direction of observation angle 0. The distributions shown are made from spline tables (Sec.~\ref{sub:spline_tables}); directionality is neglected when using the first-guess approximation of Sec.~\ref{sec:analytic_approximation}.}
\label{fig:photonics_pdf}
\end{centering}
\end{figure*}

A higher-quality but slower parametrization of $\Lambda$ is obtained using a multi-dimensional spline surface \cite{photospline} fit to the results of Monte Carlo simulations of many different source configurations. This provides a high-quality parametrization of light propagation and is, in general, used for all final analysis results.

For an approximately point-like source like an electromagnetic (EM) shower, $\Lambda$ depends on nine parameters. For the case of IceCube, however, symmetries of the detector, in particular the approximate azimuthal and lateral translational symmetry of light propagation, allow the parametrization of $\Lambda$ in terms of six: the depth and zenith angle of the source, the displacement vector connecting it to the receiver, and the difference between the time of light detection and production. Fig.~\ref{fig:photonics_pdf} shows the parameterization evaluated at a single depth and zenith angle for various receiver displacements.

The resulting tables are approximately 1 GB in size and take on order $1\, \mu$s to evaluate for a particular source-receiver configuration (much longer than the approximation described above), but are sufficiently accurate that Eq.~\ref{eq:poissonllh} can be used directly with full knowledge of light propagation in ice. Using splines also allows analytic evaluation of likelihood gradients, useful when fitting for geometric parameters (Sec.~\ref{sec:cascades}), as well as the possibility of convolving the timing distributions with additional effects such as the PMT transit time spread. The same approach, albeit with a different parametrization, can also be used to tabulate light yield from sources with other geometries, like minimum ionizing muons.

\section{Waveform Unfolding}
%\section{Photocathode Current Reconstruction}
\label{sec:photon_counting}

\begin{figure*}
\begin{centering}
\begin{subfigure}{0.48\linewidth}
	\includegraphics[width=\linewidth]{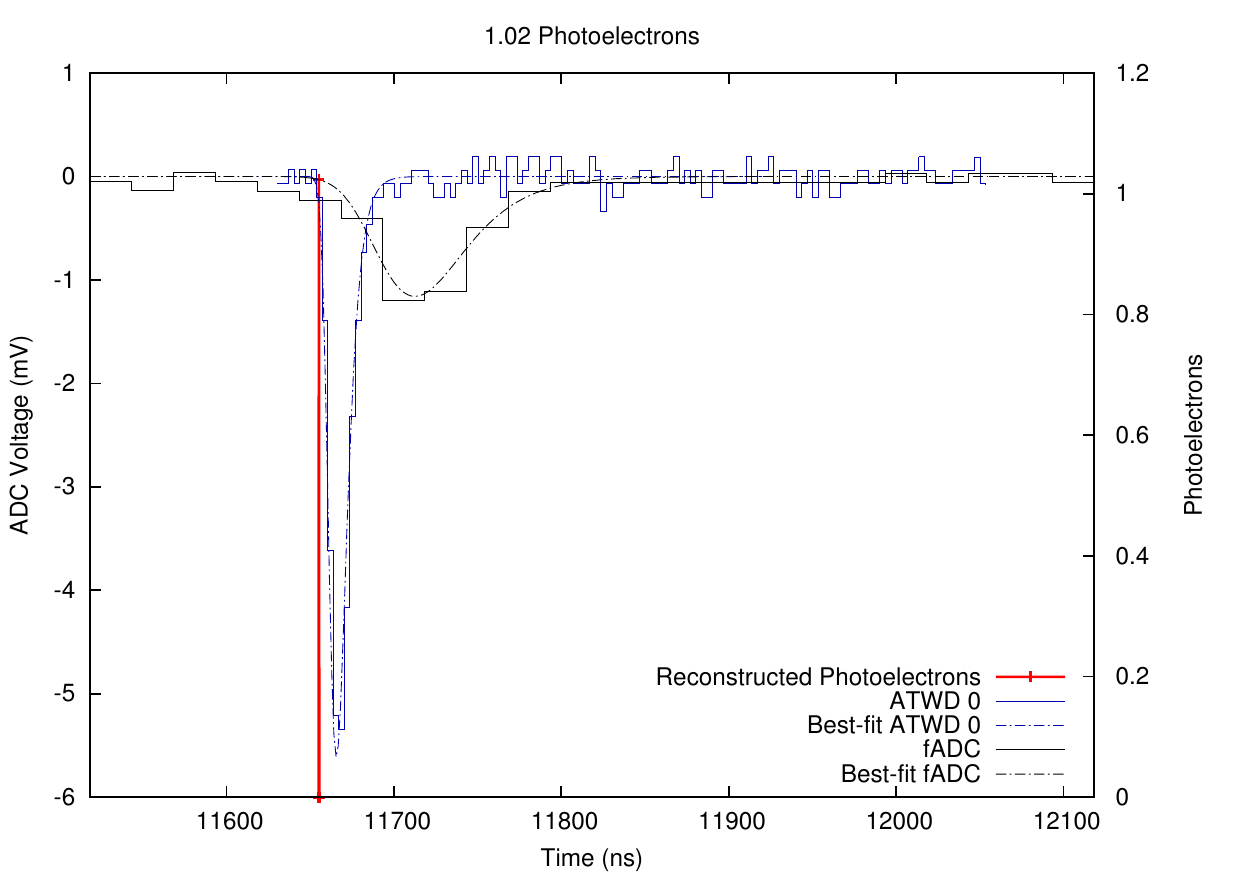}
	\caption{Unfolding of a simple waveform containing one detected photon, showing good agreement between the best-fit reconstruction and the data in all active digitizers. Both the total reconstructed amplitude (top label) and number of pulses (red line) agree with the single photon interpretation of these data.}
\end{subfigure}
\begin{subfigure}{0.48\linewidth}
	\includegraphics[width=\linewidth]{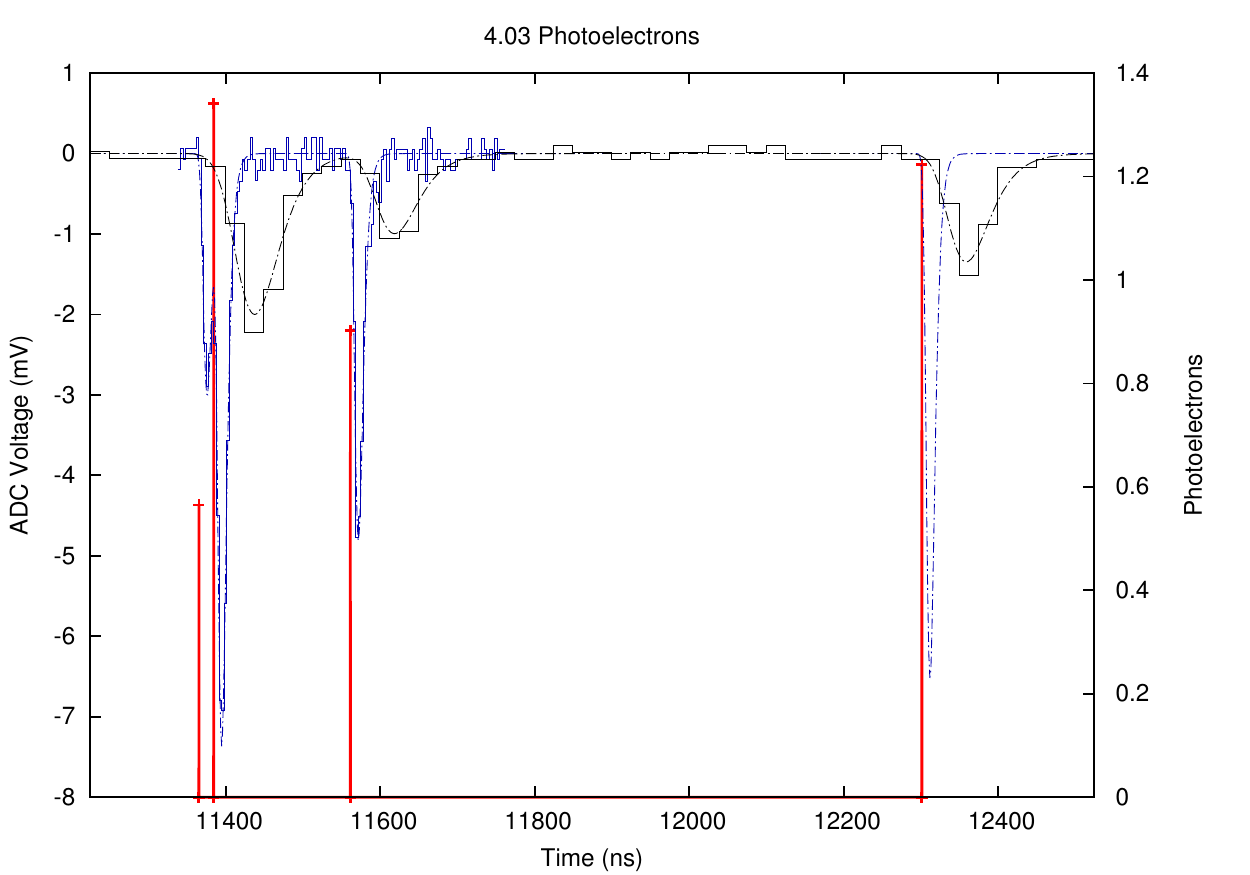}
	\caption{Unfolding a more complicated waveform, illustrating the smooth transition between digitizers and handling of pileup. The far-right pulse was recorded after the ATWD buffer was filled and is reconstructed from fADC data only. The heights of the first two pulses show the PMT amplification variance.}
\end{subfigure}
\caption[Examples of waveform unfolding]{Examples of waveform unfolding in data from the IceCube detector for both simple and complex waveforms. The lines marked {\it Best-fit} are predictions of the various digitizer read-outs given the reconstructed PMT hits. For a perfect reconstruction, and with no noise in the data, these lines would exactly match within the digitizer step (typically 0.15~mV in the highest-gain ATWD and 0.1~mV in the fADC). The vertical lines with crosses at the top represent the times and amplitudes of the unfolded pulses relative to the right-hand axis.}
\label{fig:wavedeformdemo}
\end{centering}
\end{figure*}

\begin{figure}
\begin{centering}
\includegraphics[width=\linewidth]{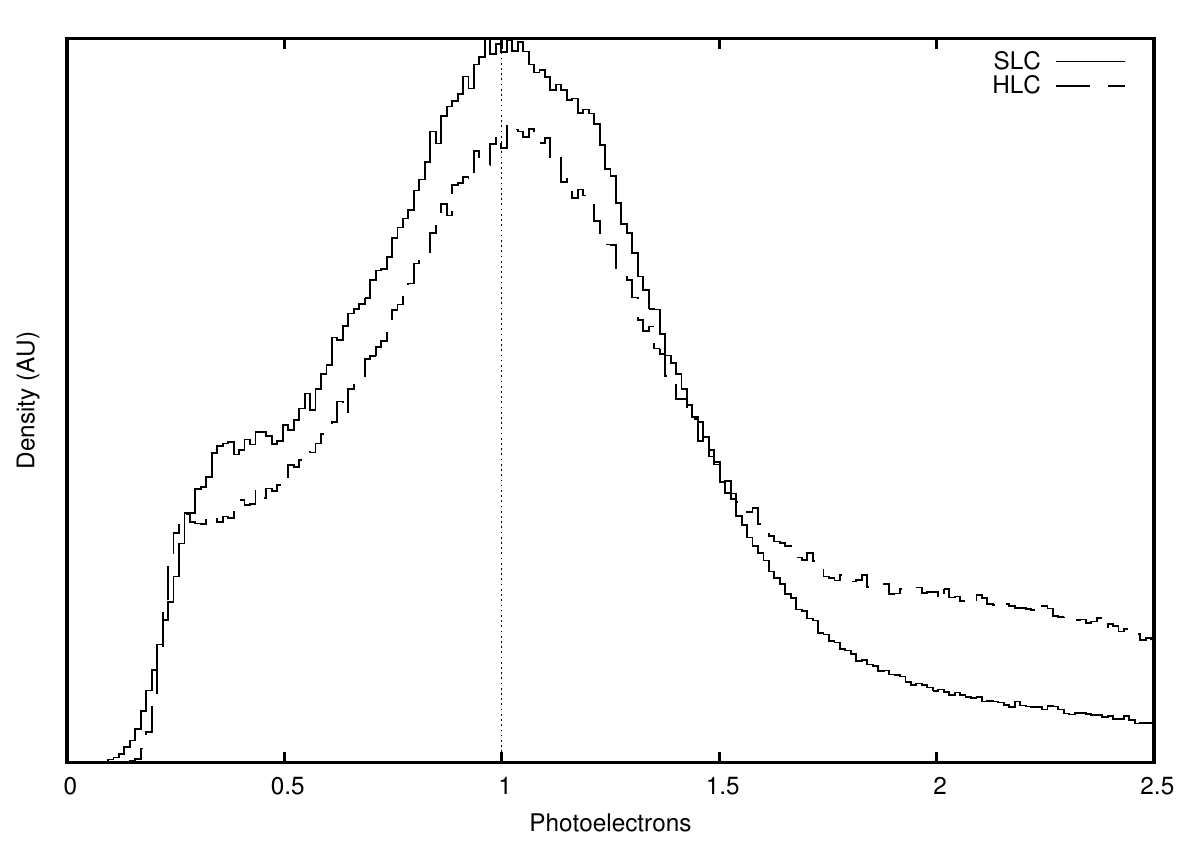}
\caption{The charge distribution of photomultiplier readouts (mostly single photoelectrons) measured in the IceCube detector. The width of this distribution arises from stochasticity in the electron cascade process in the PMT \cite{2010NIMPA.618..139A}. The 0.2~PE trigger level is clearly visible on the left. The two readout modes shown (HLC and SLC, for ``hard'' and ``soft'' local coincidence \cite{daqpaper}) are based on an in-module trigger decision; HLC hits are more likely to be physics events than noise, and so have longer readouts with more detailed waveforms. The charge resolution of both readout modes is equivalent for isolated single-photoelectron pulses; the high-charge tail in the HLC charge distribution is caused by the higher photon densities typically needed to satisfy the HLC trigger condition.}
\label{fig:wavedeformspedist}
\end{centering}
\end{figure}

IceCube uses waveform-recording digitizers to collect data from photomultiplier tubes (PMTs) \cite{daqpaper}. For these data to be used in Eq.~\ref{eq:poissonllh}, these waveforms must be transformed to a reconstructed number of photons per unit time. The PMT output is AC-coupled by a toroid transformer through a set of pulse-shaping amplifiers to four digitizers: three high-rate short-duration custom modules (Analog Transient Waveform Digitizers or ATWDs) recording the first 420~nanoseconds of the waveform at three gains with a typical sampling period of 3.3~ns and a continuous pipelined digitizer (the fast ADC or fADC) with a 25~ns sampling period. The shape applied by the amplifiers is much wider than the intrinsic width of the PMT pulse. The resulting recorded waveforms (Fig. \ref{fig:wavedeformdemo}) are then a linear combination of the characteristic shaping functions of the amplifiers with timing and amplitude related to the charge collected at the PMT anode. The unit used for this collected charge, the photoelectron (PE), is defined as the most likely deposited charge from a single photon, $\sim$1.6 picocoulombs at the typical IceCube PMT gain of $10^7$.
Note that, due to the shape of the PMT charge response function \cite{2010NIMPA.618..139A} and our discriminator settings, the mean deposited charge of triggering photons is $14\%$ lower than the most likely value.

\begin{figure*}
\begin{centering}
\includegraphics[width=\linewidth]{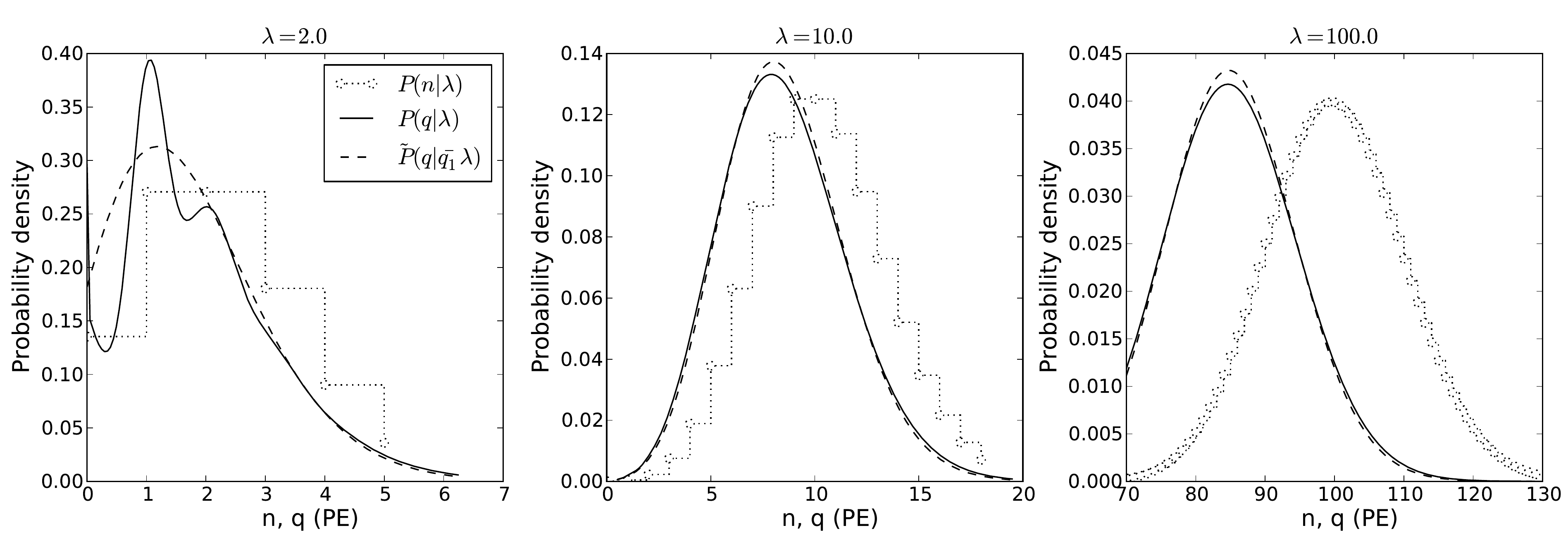}
\caption{An approximation to the distribution of charge responses given a mean number of detected photons ($\lambda$) at the PMT photocathode. The stepped line shows the Poisson probability of ejecting $n$ electrons from the photocathode, given $\lambda$. The solid line shows the numerical convolution of this Poisson distribution with the charge response function (Fig.~\ref{fig:wavedeformspedist}), yielding the given number of photoelectrons (PE) at the anode. The dashed line shows the analytic approximation we use to this convolution, using the extension of the Poisson probability distribution to real numbers $q$. The shift in the peak centers between the stepped and dashed/solid lines (between $n$ and $q$) is related to our choice of the definition of $\lambda$ with respect to the PMT quantum efficiency and the shape of the charge response functions of the PMTs \cite{2010NIMPA.618..139A}.}
\label{fig:poissonvalidity}
\end{centering}
\end{figure*}

A non-negative linear simultaneous unfolding of all digitizers (in this case, using the Lawson-Hanson NNLS algorithm \cite{Lawson:1974}) can then be applied using the shaping functions as a basis to recover the collected PMT charge as a function of time (Fig. \ref{fig:wavedeformdemo}). Charge resolution obtained by this method, which is dominated by the width of the charge response of the Hamamatsu R7081 photomultiplier \cite{2010NIMPA.618..139A,doublechooz_calib}, is typically around 30\% at the single photon level (Fig. \ref{fig:wavedeformspedist}). As photon statistics accumulate (Sec. \ref{sec:standardcandle}), this uncertainty is reduced by averaging over many electron cascades, and PMT charge counting is dominated by purely Poissonian effects, as in Eq.~\ref{eq:energyrecownoise}. For our purposes, it is sufficient to approximate the charge resolution by allowing the photon count ($k$) in Eq.~\ref{eq:poissonllh} to take non-integer values, replacing the $k!$ normalization term with $\Gamma(k+1)$. At low amplitudes, where only a few electron cascades contribute, the contribution of the single photoelectron (SPE) width is maximal (30\%), but the Poisson uncertainties on photon collection remain larger. As a result, our pure-Poissonian approximation can be used for the distribution of collected charges at all amplitudes (Fig. \ref{fig:poissonvalidity}).

Relative timing from the electronics and unfolding is typically accurate to around 1~ns for consecutive non-overlapping pulses on a single DOM while the ATWD is active. This is comparable to the transit time spread of the PMT and degrades to 8~ns when only the fADC data are available. Pulse timing in complicated waveforms can be substantially more uncertain (up to 10~ns for the ATWD), but still correctly reproduces the timing distribution of photon cascades at the PMT anode even at very high total amplitudes (Sec. \ref{sec:standardcandle}). This allows the use of a Poisson likelihood \eqref{eq:poissonllh} for particle reconstruction in events at all energies.

\section{Energy Scale Calibration}

The energy reconstruction capabilities presented here require linear behavior of the photomultiplier tubes and readout electronics over several orders of magnitude in collected charge. Precise understanding of light propagation in the ice with horizontal instrumentation spacing similar to the average absorption length of light ($\sim125$ m) is also required, as are verification of purely Poissonian fluctuations in collected charge as in Eq.~\ref{eq:poissonllh} and calibration of the absolute energy scale of the detector. Due to the nature of the IceCube detector as a naturally occurring volume of ice, the verification of these properties of the detector must be conducted in situ.

Calibration of the IceCube detector relies on built-in reference electronics, calibration LEDs integrated into each DOM, in-ice calibration lasers, observations of dust concentrations and optical properties in the side walls of ice boreholes taken during IceCube construction \cite{dust_logger}, as well as observed physics data. The modeling of light propagation in ice \cite{aha,spice} is conducted using the LED and sidewall dust observations. Final energy scale calibration uses minimum-ionizing muons as a standard brightness candle in order to probe the detector volume with Cherenkov light. Calibration laser data is then used to establish the linearity of the detector's light response and thereby the energy ladder used to reconstruct high-energy particles.

\subsection{Minimum-Ionizing Muons}
\label{sec:energyscale}

% Jake F.

Minimum-ionizing muons, created in atmospheric cosmic-ray showers, are well-suited for energy scale calibration because they have constant known light emission, are abundant, and leave well-defined tracks in the detector.  Obtaining a large sample of $\sim 100\,\,\unit{GeV}$ single muons, their positions and directions are reconstructed to high precision. The observed PMT charges are then compared to expected values, similar to the energy reconstruction procedure in Eq.~\ref{eq:poissonllh}.

To isolate a sample of events we first make cuts on track quality parameters, including the number of on-time PMT hits and the fit quality of the track reconstruction.  We then select low-energy single muons by searching for tracks that deposit little light in the outer strings and appear to stop in the detector fiducial volume.  Finally, we require the tracks to be inclined $45^\circ$-$70^\circ$ with respect to the straight downgoing direction to ensure the muon's Cherenkov cone is incident on the active side of our PMTs, which face down towards the bottom of the glacier. These cuts provide a sample of 70,000 events in 30 days of data taken with IceCube in its 79-string configuration.  Monte Carlo studies of the detector response to cosmic ray air showers \cite{CORSIKA} show this sample consists of $> 95\%$ single muons with a median energy of 82 GeV at the detector center.  Distributions of event observables show good agreement between simulation and experimental data.

\begin{figure}
\begin{centering}
\includegraphics[width=\linewidth]{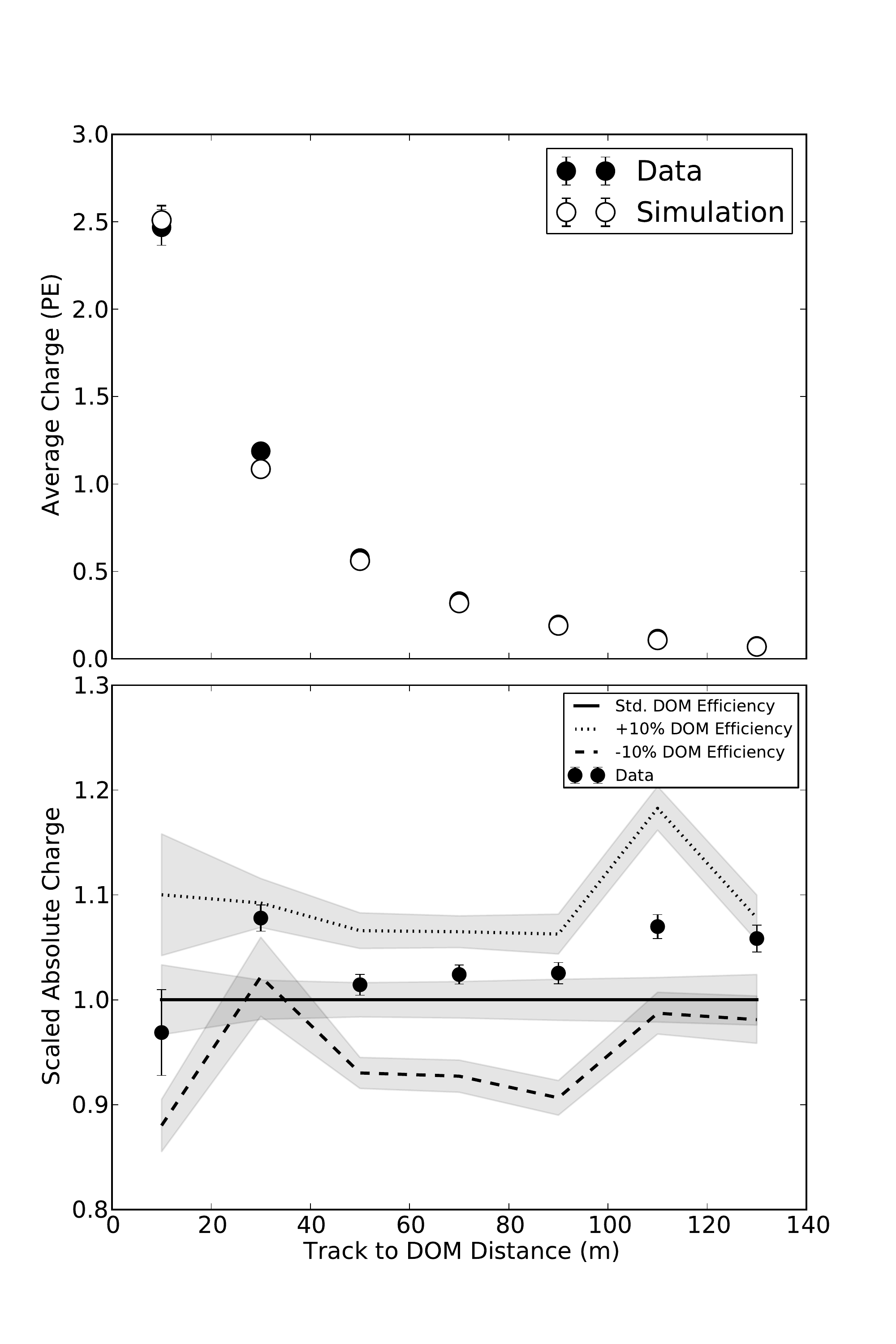}
\caption[Absolute charge measurements with minimum-ionizing muons]{Absolute charge measurements with minimum-ionizing muons.  Top: average observed charge vs. distance from the DOM to the reconstructed muon track, shown for both data and simulation.  Due to the selection of minimum-ionizing, single muons, the observations are dominated by single photoelectrons.  Bottom: the average charge vs. track-DOM distance normalized to the charge expected from standard IceCube simulation.  Muon data and simulation are shown.  As a proxy for an altered energy scale, simulations with altered DOM efficiencies are shown for comparison.  The observed charge is slightly higher than nominal values but below the charge expected for a DOM efficiency increased by 10\% (upper band). Error ranges reflect statistical uncertainties on the shown sample only without including statistical uncertainties in the simulation dataset to which they were normalized.
}
\label{fig:abssens}
\end{centering}
\end{figure}

To calibrate the energy scale, we focus on a subset of IceCube DOMs in the deep part of the detector where the glacial ice is exceptionally clear (absorption lengths of $\sim 200\,\,\unit{m}$ \cite{spice,aha}).  For each DOM in this region, we first reconstruct the muon track while excluding all information from the DOM in question.  Simulation studies show this procedure successfully reconstructs the muon direction and the track-DOM distance within $\sim 2^\circ$ and $\sim 10 \,\,\unit{m}$ of the true direction and position. After binning the observed charge based on the track-DOM distance, we find a $\lesssim 5\%$ average excess of charge in data compared to nominal values (Fig.~\ref{fig:abssens}) with up to 9\% deviations at certain distances.

%to do:
%-update plots
%-reference PMT paper?
%-change some #'s?  5.1%? 8%?
%-some text can be cut or moved to figure captions

\subsection{High-energy Linearity Calibration}
\label{sec:standardcandle}

% Jakob

The IceCube detector includes two 337-nm pulsed nitrogen lasers \cite{standard_candle_icrc} that are operated through adjustable optical attenuators to produce pulses of light that are identical except for the number of emitted photons and correspond approximately to the light output of electromagnetic showers in the 1-100 PeV range. This strictly linear behavior can be used to verify the linearity of the DOM electronics and the photon-counting procedure (Sec.~\ref{sec:photon_counting}), extending the energy scale established at low photon counts with minimum-ionizing muons to the regime where the DOM collects thousands of photons.

\begin{figure}
\begin{centering}
\includegraphics[width=\linewidth]{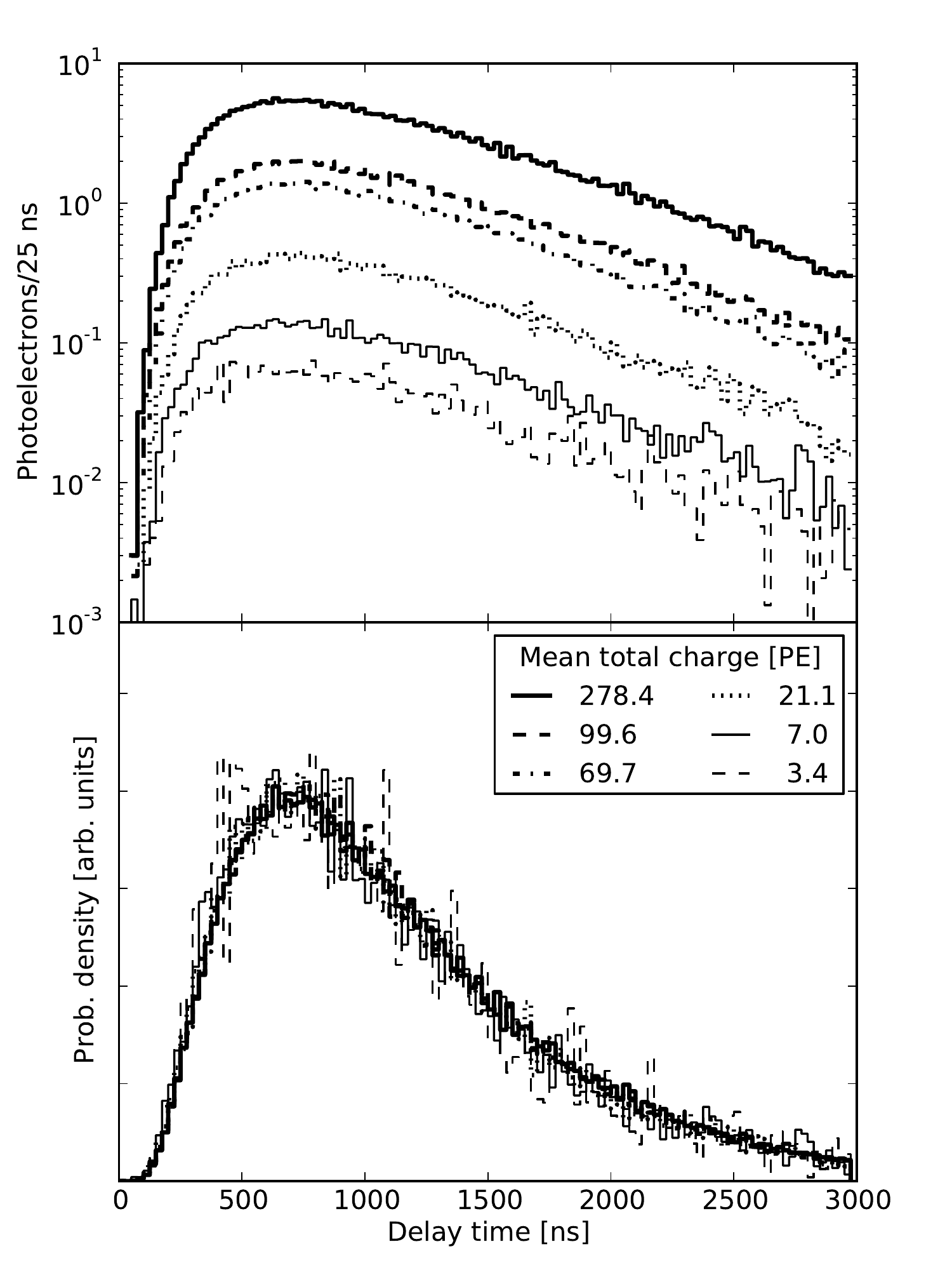}
\caption{Top: charge collected at a single DOM as a function of time in response to calibration laser flashes with 6 different transmittance settings at a distance of 246 m from the laser. Bottom: the same distributions as the top, but normalized to each transmittance setting. The charge distributions are scaled copies of each other, whether derived from isolated pulses (lowest dashed line) or high charge bunches (highest solid line).}
\label{fig:sc_waveform_scaling}
\end{centering}
\end{figure}

\begin{figure}
\begin{centering}
\begin{subfigure}{\linewidth}
	\includegraphics[width=\linewidth]{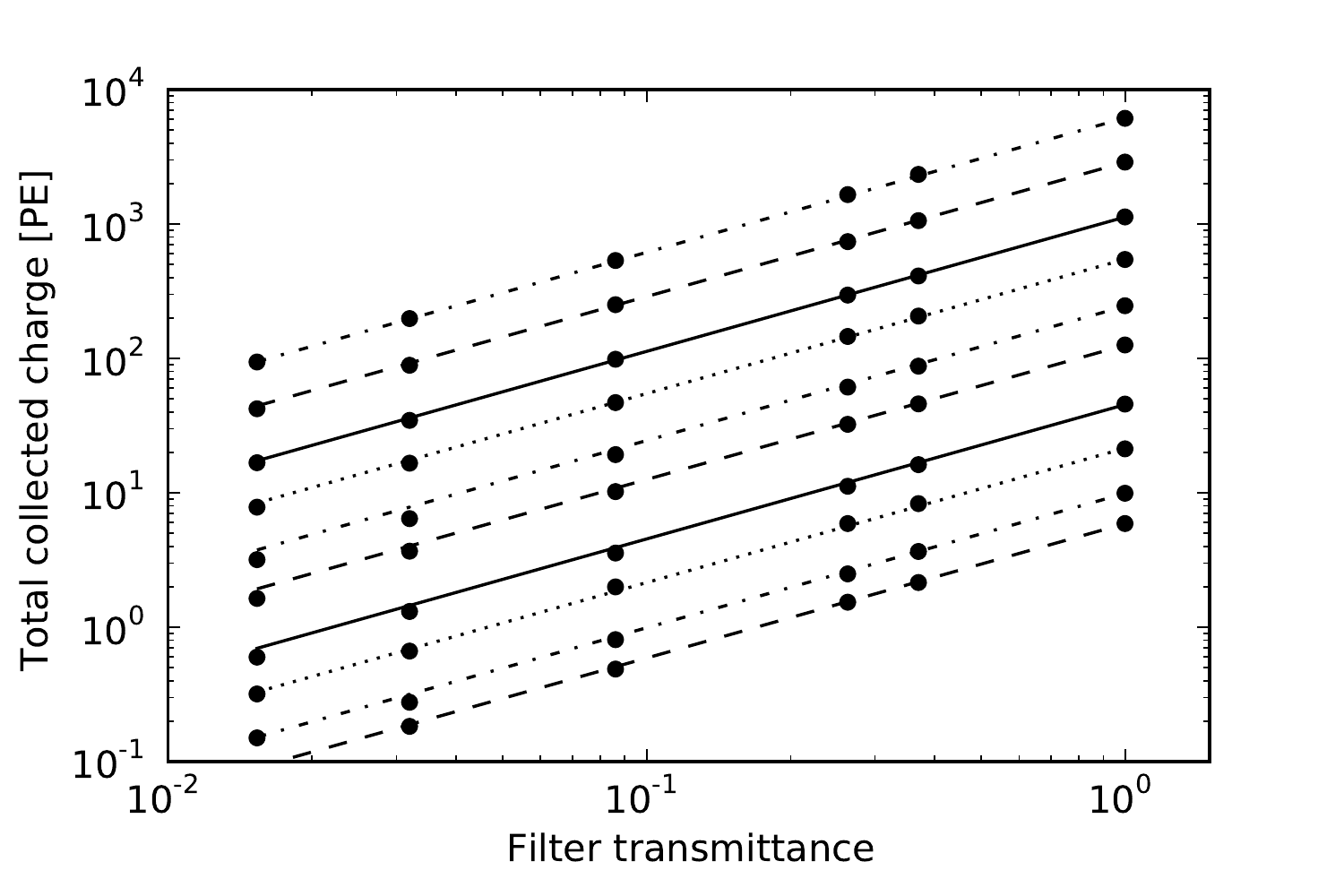}
	\caption{Charge collected in ten typical DOMs as a function of filter transmittance. The observed PMT charge shown on the vertical axis is proportional to the number of collected photons, which is in turn proportional to the filter transmittance shown on the horizontal axis. Differences between DOMs are due to different distances from the laser calibration source.}
	\label{fig:standardcandle-scaling}
\end{subfigure}
\begin{subfigure}{\linewidth}
	\includegraphics[width=\linewidth]{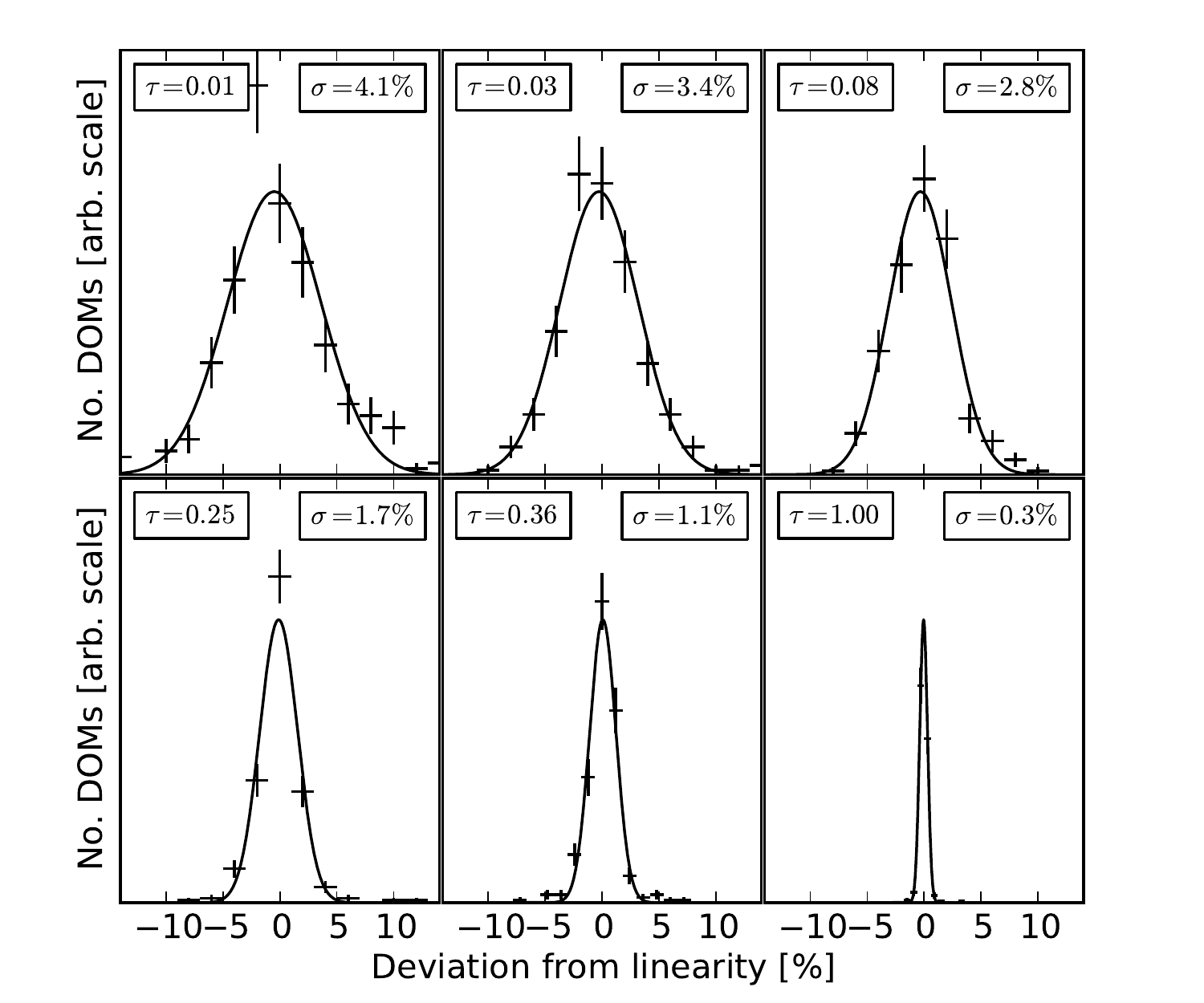}
	\caption{Fractional deviations of observed light amplitudes from predictions, assuming a linear scaling, in 276 DOMs that triggered in response to calibration laser pulses at six different filter transmittance settings ($\tau$ in each panel). Each entry in each histogram is the proportional deviation of the collected charge from a best-fit line like those shown in (\subref{fig:standardcandle-scaling}). The linear fit effectively pivots around the highest transmittance setting (lower right panel). The largest deviations are at the lowest transmittance setting (upper left panel), and are on the scale of 5\%.}
	\label{fig:standardcandle-correlation}
\end{subfigure}
\caption[Linearity measurements with the calibration laser]{Linearity measurements using the calibration laser source. The PMT, digitization electronics, and photon reconstruction procedure (discussed in Section~\ref{sec:photon_counting}) respond linearly to photon fluxes spanning 4 orders of magnitude.}
\label{fig:standardcandle-linearity}
\end{centering}
\end{figure}

If the DOM response is linear then the time distributions of PMT charge in response to laser pulses with different attenuation settings must be scaled copies of each other (Fig.~\ref{fig:sc_waveform_scaling}). Further, the total charge collected over a given time window must be proportional to the intensity of the pulse. Since the pulses trigger DOMs both close to the laser and those hundreds of meters away, the linearity of the DOM response both in charge and timing structure can be established from fractions of to many thousands of PE per recorded waveform (Fig.~\ref{fig:standardcandle-linearity}). Any non-linearity in either the hardware or photon reconstruction would appear in two possible ways: as a distortion of the total charge vs. attenuator setting (see Fig.~\ref{fig:standardcandle-linearity}), and as distortions in the shapes of the waveforms at different amplitudes as a result of different instantaneous photon amplitudes at different points in the waveform.

\begin{figure}
\begin{centering}
\includegraphics[width=\linewidth]{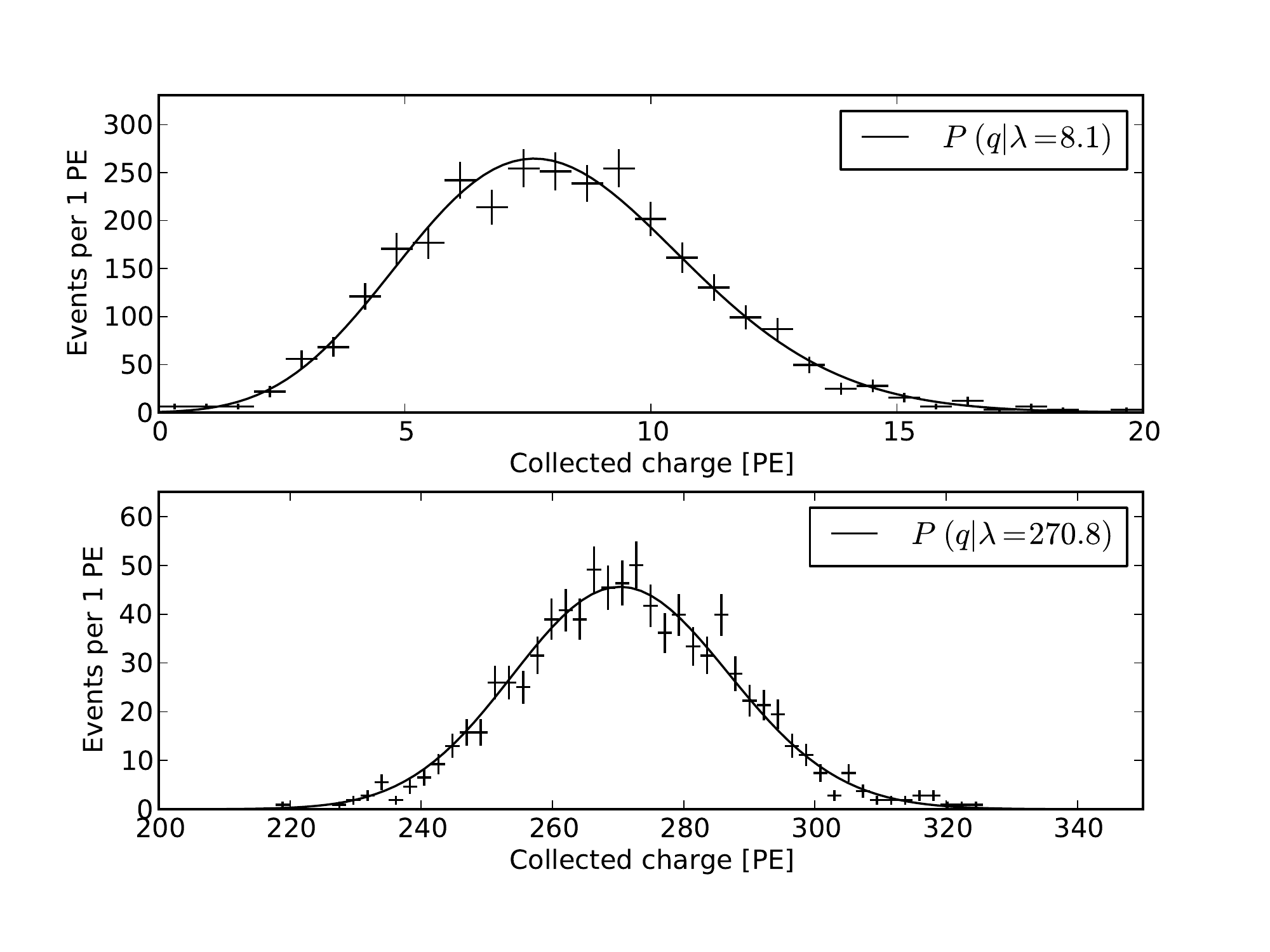}
\caption{
Total charge collected in response to 1882 calibration laser flashes at maximum transmittance in two DOMs at distances of 297~m (top panel) and 194~m (bottom panel). In both cases, the fluctuations in collected charge follow the predicted distribution from Eq.~\ref{eq:poissonllh}.
}
\label{fig:sc_waveform_fluctuations}
\end{centering}
\end{figure}

In addition to the linearity of the waveforms, and the ability to interpret waveforms containing many photoelectrons statistically as in Eq.~\ref{eq:poissonllh}, the repeatability of the calibration laser pulses allows tests of the statistical uncertainty in light collection. At levels above the 0.2 photoelectron discriminator \cite{daqpaper}, the distribution of measured charges in the DOMs is well-described by a Poisson distribution (Fig. \ref{fig:sc_waveform_fluctuations}), verifying the likelihood model used in the reconstructions \eqref{eq:poissonllh}.

\section{Cascades}
\label{sec:cascades}

% Jakob van Santen

Reconstruction of the energies of the isolated cascades produced in $\nu_e$ and neutral-current events is the simplest scenario. As the light deposition pattern for such events is independent of energy on the scale of IceCube instrumentation (Fig.~\ref{fig:em_longitudinal_profile}), the likelihood model and scaling formulae \eqref{eq:poissonllh} are obeyed exactly for cascade events. Thus, the energy of a cascade with a known orientation and vertex can be recovered using a template cascade by numerical maximization of Eq.~\ref{eq:energyrecownoise}. The likelihood $\mathcal{L}(E)$ \eqref{eq:energyrecownoise} is strictly concave, with a single maximum, and can be optimized easily with any standard algorithm; here we use the Non-monotonic Maximum Likelihood algorithm (NMML) \cite{nmml} as in Sec.~\ref{sec:millipede}.

The deposited energy from such neutrino interactions is nearly identical to the neutrino energy for charged-current $\nu_e$ interactions (Fig.~\ref{fig:nue_visible_energy}). For neutral-current interactions (all flavors), much larger variations are possible due to the unseen outgoing neutrino and the reconstructed deposited energy is a lower limit on the neutrino energy. This is true both because of the potentially large amount of missing energy in the outgoing neutrino in neutral current interactions and due to the $\sim 15\%$ lower light yields typical of hadronic showers \cite{wiebusch_thesis}. IceCube is not capable of resolving the differences between these event types and all quoted cascade energies, which are given as the reconstructed deposited energy using an EM shower template, are therefore lower limits on the energies of the neutrinos that produced them.

\begin{figure}
\begin{centering}
\begin{subfigure}{\linewidth}
\includegraphics[width=\linewidth]{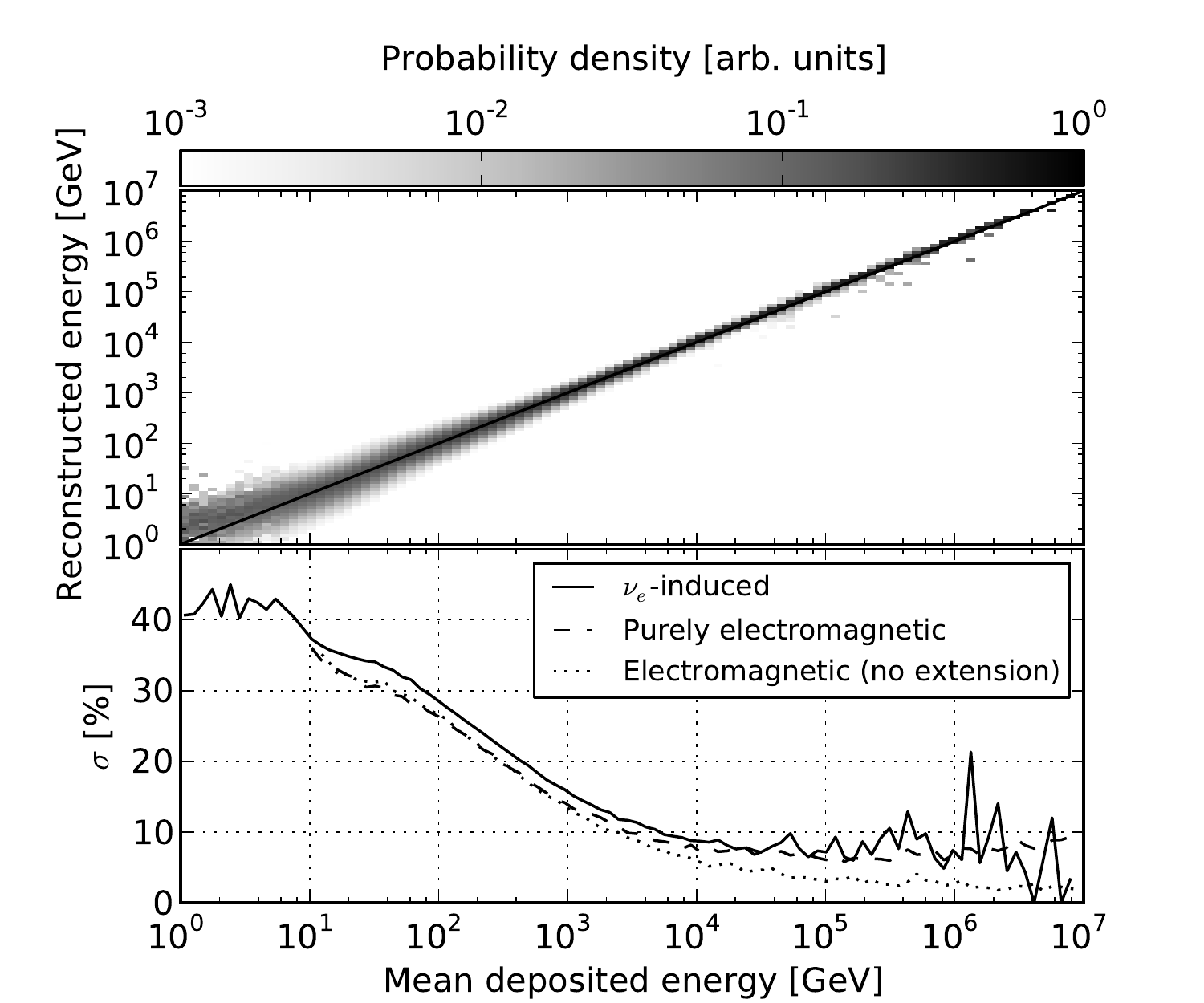}
\subcaption{Reconstructed energy deposition of cascade events as a function of true energy deposition.}
\label{fig:monopod_performance:2d}
\end{subfigure}
\begin{subfigure}{\linewidth}
\includegraphics[width=\linewidth]{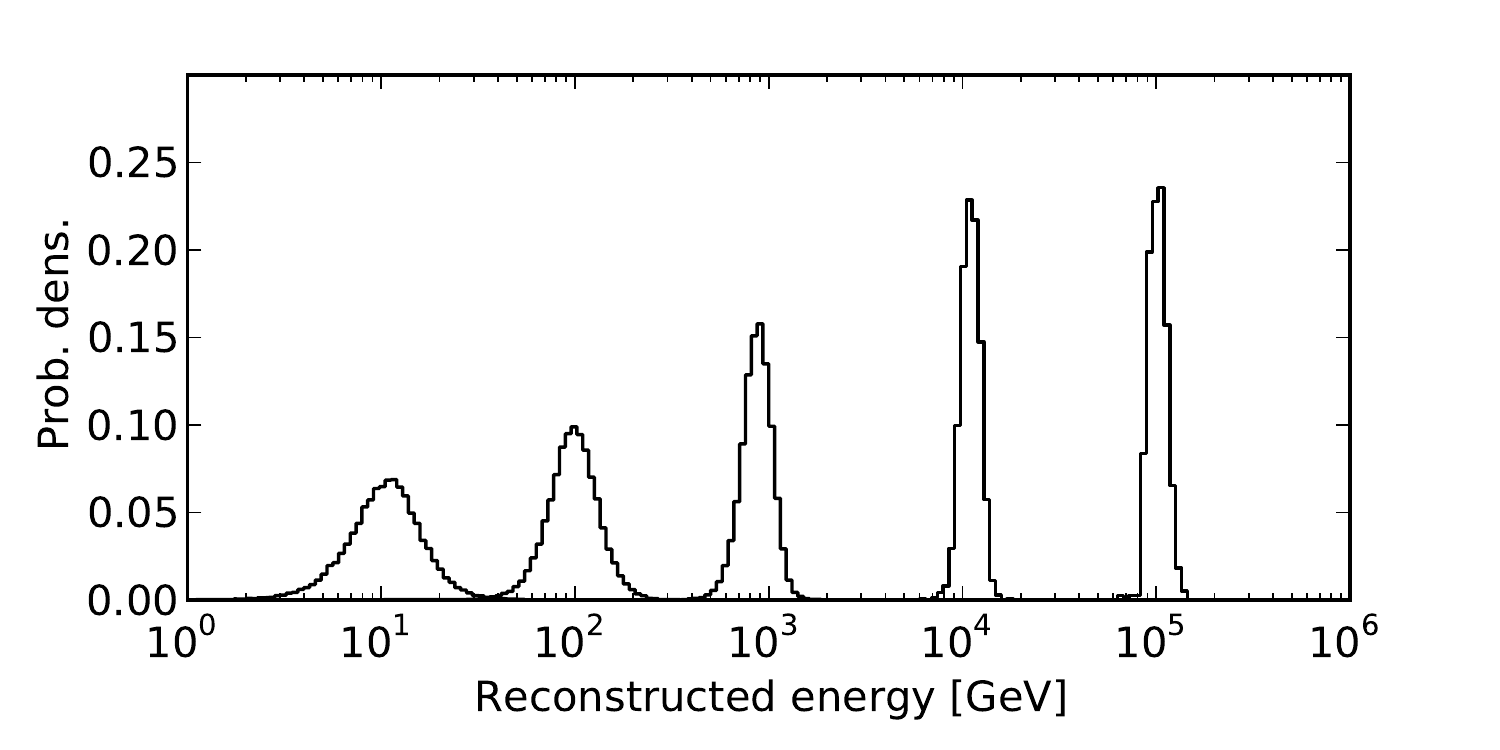}
\subcaption{Slices of the reconstructed-energy distribution shown in (\subref{fig:monopod_performance:2d}) at fixed true total energy depositions of $10^1$, $10^2$, $10^3$, $10^4$, and $10^5$ GeV.}
\label{fig:monopod_performance:slices}
\end{subfigure}
\caption[Energy resolution for $\nu_e$]{
Performance of cascade energy reconstruction on simulated $\nu_e$ events with known vertices and directions throughout the IceCube array. The horizontal axis in (\subref{fig:monopod_performance:2d}) shows the mean electromagnetic-equivalent deposited energy (correcting any hadronic component of the showers to have the energy of an EM shower with the same light yield). The inherent variance in the energy fraction in the hadronic energy in the neutrino interaction contributes to the reconstruction uncertainty shown in the lower panel of (\subref{fig:monopod_performance:2d}) (solid line) relative to purely electromagnetic showers (dashed line).
Fluctuations in the light yield and charged particle content of hadronic showers also increase the uncertainty in the energy measurement at all energies (difference between dashed and solid line). Above 100 TeV, the resolution of the single-cascade template method is limited by the unmodeled longitudinal extension of the showers (dashed vs. dotted lines).
}
\label{fig:monopod_performance}
\end{centering}
\end{figure}

The deposited energy resolution for contained $\nu_e$ events is dominated by statistical fluctuations in the collected charge at low energies, improving from 30\% at 100~GeV to 8\% at 100~TeV, at which point the extension of the shower begins to distort the reconstruction (see Fig.~\ref{fig:monopod_performance}). This high-energy limit on resolution is similar to uncertainties in the modeling of scattering and absorption in the ice sheet \cite{spice,aha}, which contribute a 10\% systematic uncertainty to the energy.
Since we do not perform analysis-level event selection in this article, events shown in Fig.~\ref{fig:monopod_performance} are distributed across the entire detector.
Below 100 GeV, events outside the densely-instrumented DeepCore subarray \cite{DeepCore_design} typically only deposit a few photons, resulting in the figure reflecting generic IceCube performance rather than that typical for a DeepCore-specific low energy analysis that would exclude these few-photon events.

\begin{figure}
\includegraphics[width=\linewidth]{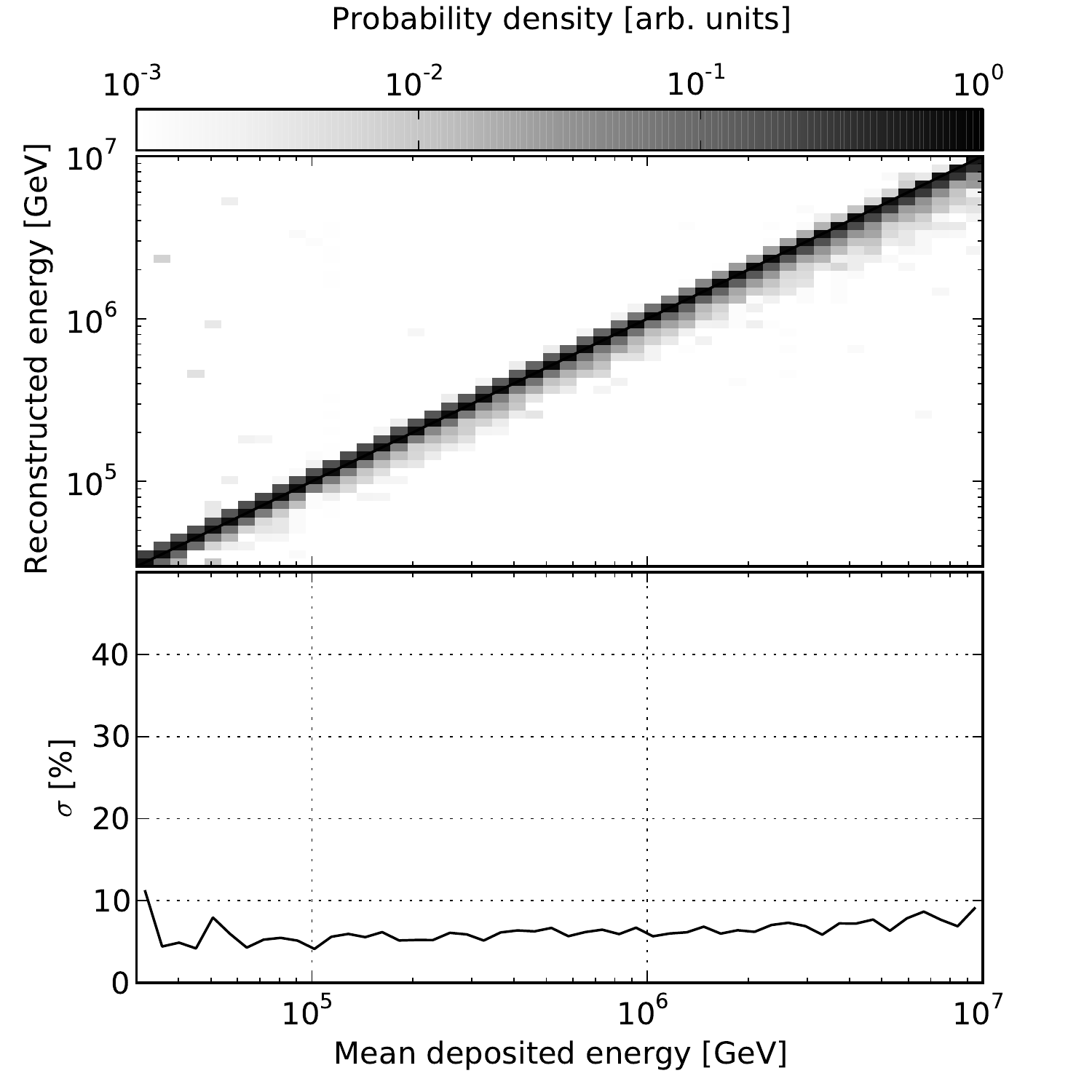}\\
\includegraphics[width=\linewidth]{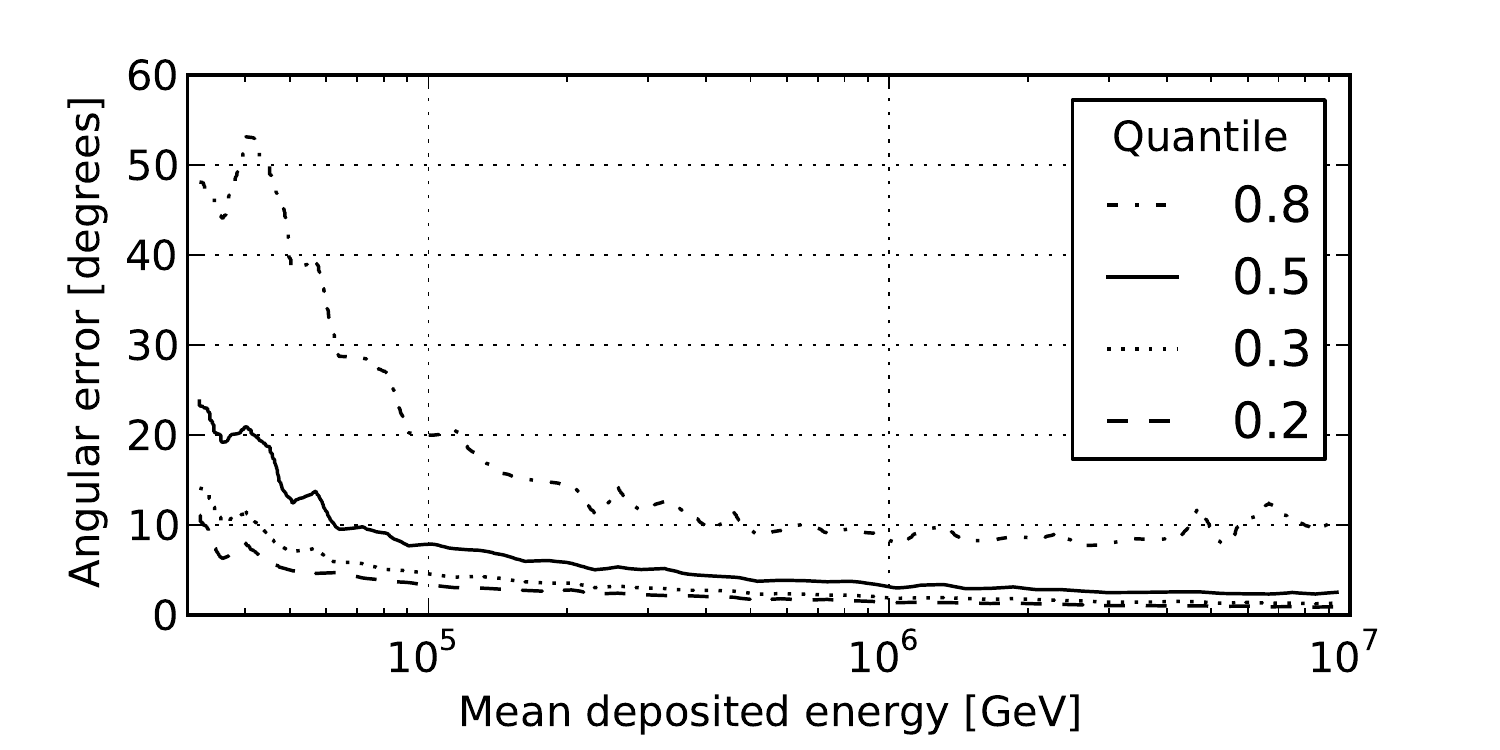}
\caption{Reconstruction performance for $\nu_e$ events in deposited energy (top) and direction (bottom) at final level for a recent IceCube analysis searching for astrophysical neutrino events at typical energies of 100 TeV \cite{hese_paper}. Energy resolution includes the effects of the event selection in the analysis as well as the effects of uncertainty induced by the vertex and angular reconstruction (bottom). The results are generally similar to those shown over the equivalent energies in Fig.~\ref{fig:monopod_performance}, although with a larger population of outliers due to limited vertex resolution. Event selection effects, along with reduced photon statistics, worsen angular resolution at the left near the lower energy threshold of the analysis.}
\label{fig:hese_resolutions}
\end{figure}

The light-yield templates ($\Lambda$) in Eq.~\ref{eq:poissonllh} are independent of energy but are functions of the event topology, in particular the neutrino interaction vertex position and shower orientation (Fig.~\ref{fig:photonics_pdf}). Simultaneous maximization of the likelihood in $E$, direction, and vertex position, through the influence of the last two on $\Lambda$, allows reconstruction of these parameters as well. In this full-reconstruction case, energy resolution is very similar (see Fig.~\ref{fig:hese_resolutions}) and a systematics-dominated angular resolution on the order of $15^\circ$ is achieved for energies of $\gtrsim 100$ TeV.

Computational performance of cascade reconstruction is greatly enhanced by using a standard numerical minimizer for the topological parameters and then employing a second internal minimization algorithm to solve for the best-fit value of $E$ at each iteration. This exploits the relatively long time required to evaluate $\Lambda$ relative to the multiplication $\Lambda E$ as well as the fact that the sub-problem of energy reconstruction is nearly linear. Since solving for $E$, given $\Lambda$, requires less CPU time than evaluating $\Lambda$, which is constant with energy, this procedure reduces by 1 the effective dimensionality of the problem. Methods and performance for angular and positional reconstruction of EM and hadronic showers will be addressed in more detail in a future publication.

\section{Tracks: Muons and Taus}

% Nathan, Gary?

Measurement of the energies of particles producing throughgoing tracks is more complicated than in the case of cascades. At low energies ($\lesssim 100\,\,\unit{GeV}$), the range of muons in ice is short enough that all muon energy is deposited in the detector and a calorimetric approach can be taken. At the higher energies on which this paper is focused, muons typically have a range longer than the length of the detector. This presents two immediate complications. First, the muon's point of origin (the neutrino vertex for muons produced in $\nu_\mu$ interactions) is unknown for events starting outside the detector and so a measurement of muon energy at the detector can provide only a lower bound on the muon's energy at production, the quantity of interest for reconstructing muon neutrino energies. Second, the muon energy must be estimated only from the properties of the light emitted by the muon during the portion of its track in the detector, in particular the differential energy loss rate ($dE/dx$).

Above the minimum-ionizing regime ($\gtrsim 1$ TeV, typical for IceCube), the average energy loss rate of muons increases approximately linearly with energy \cite{pdg}. Energy loss in such events is dominated by stochastic processes such as bremsstrahlung, photonuclear interactions, and pair production involving rare exchanges of high-energy photons \cite{pdg}. The result is that large fluctuations are possible in the amount of energy distributed within IceCube even for muons of the same energy.

Not all particles of interest are through-going muons: starting muons and taus have energy loss patterns that deviate substantially from the ideal case of constant average energy loss. Muon neutrinos that undergo charged-current interactions inside the instrumented volume have no energy losses before the interaction vertex, followed by a hadronic cascade at the interaction vertex and stochastic losses along the outgoing muon track. Charged-current $\nu_{\tau}$ interactions above $\sim 1\,\,\unit{PeV}$ can yield taus that travel a detectable distance before decaying, producing a hadronic cascade followed by a track of direct electromagnetic losses from the $\tau$ itself, in turn followed by a cascade or muon from the $\tau$ decay.

\section{Muon Energy Loss Reconstruction}
\label{sec:muons}

\begin{figure}
\begin{centering}
\begin{subfigure}{\linewidth}
\includegraphics[width=\linewidth]{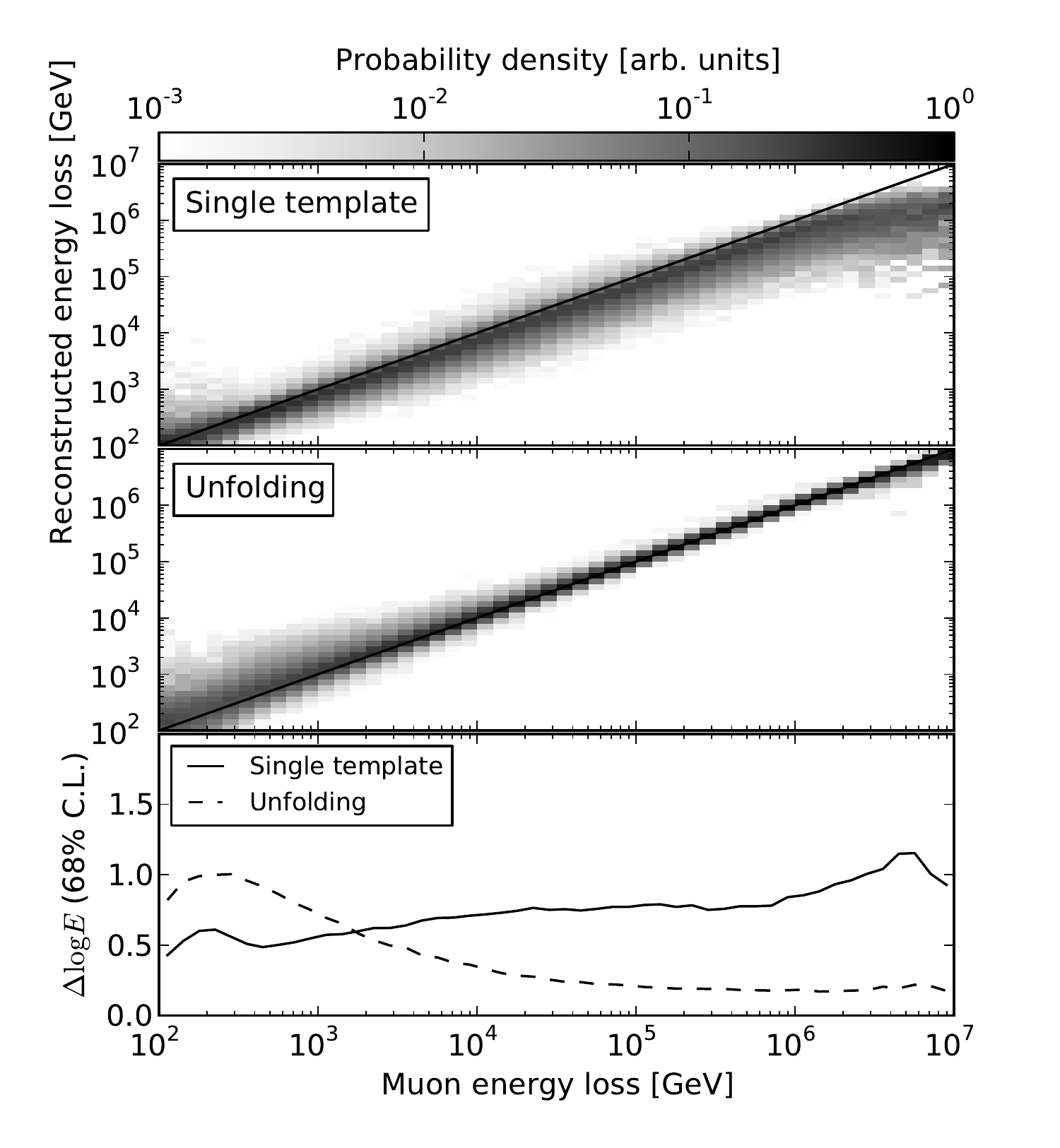}
\caption{Total energy loss reconstructed by the single-template method analogous to the cascade reconstruction method of Sec.~\ref{sec:cascades} (top panel) and by the unfolding method of Sec.~\ref{sec:millipede} (middle panel). The bottom panel shows, for each energy-loss bin, the $1 \sigma$ range of energy losses reconstructed for events in that bin.}
\label{fig:dE_v_dE_monopod_v_millipede}
\end{subfigure}
\begin{subfigure}{\linewidth}
\includegraphics[width=\linewidth]{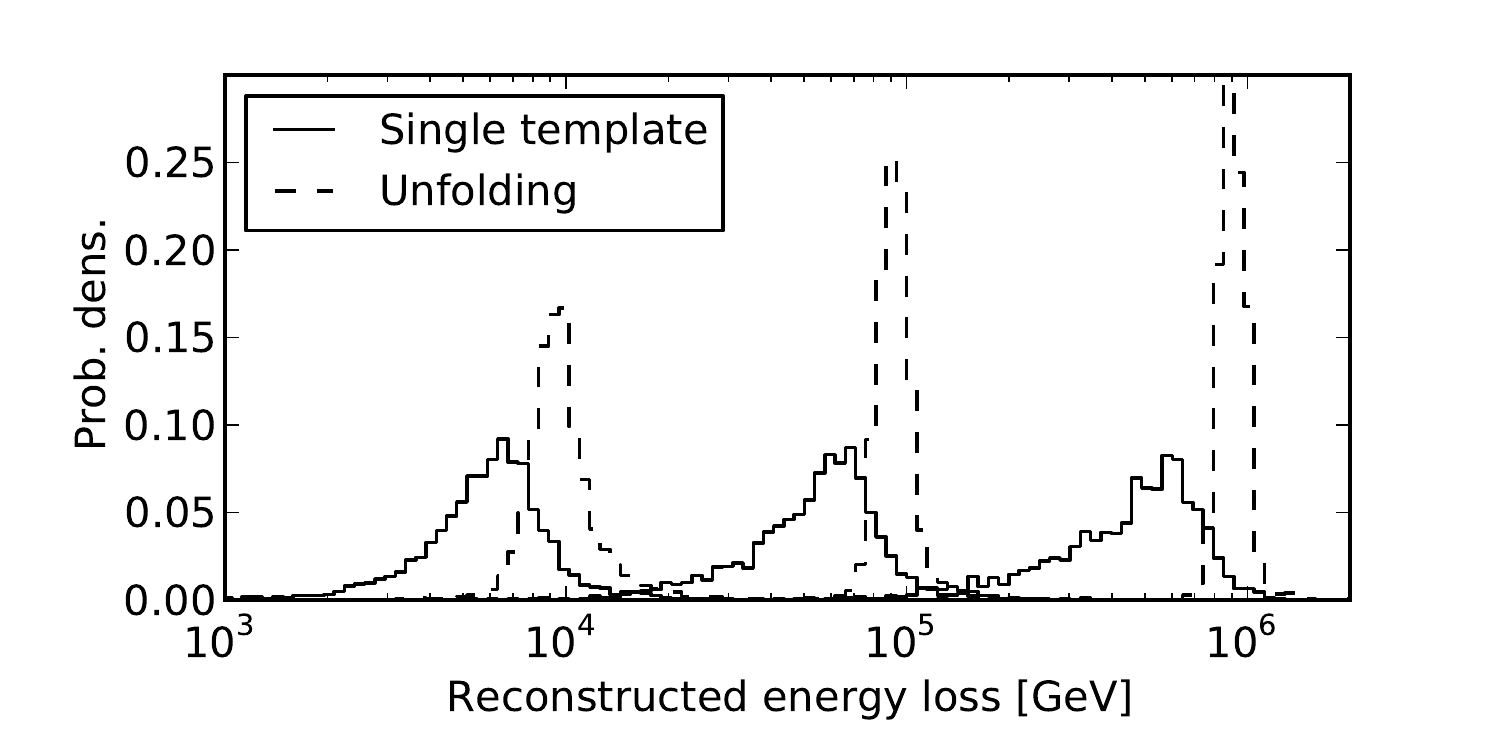}
\caption{Slices of the distribution of reconstructed total energy loss shown in (\subref{fig:dE_v_dE_monopod_v_millipede}) at fixed true total energy loss of $10^4$, $10^5$, and $10^6$ \unit{GeV}.}
\label{fig:dE_v_dE_monopod_v_millipede_slices}
\end{subfigure}
\caption{Total energy loss within IceCube from simulated single muons passing through the detector reconstructed by the single-template method, analogous to the cascade reconstruction method of Sec.~\ref{sec:cascades}, and by the unfolding method of Sec.~\ref{sec:millipede}. The continuous-loss approximation inherent in the single-muon template method becomes steadily worse as the energy deposition increases, whereas the resolution of the unfolding method improves as photon statistics accumulate, reaching a full width of 20\% above a deposition of 1~PeV.}
\end{centering}
\end{figure}

The simplest approach to reconstruction of muon energy losses is analogous to reconstruction of cascade energies: finding the best-fit scaling of a template muon to the observed light deposition. The significant event-to-event variation in muon topologies from large stochastic losses along the track causes bias and poor resolution (Fig.~\ref{fig:dE_v_dE_monopod_v_millipede}, \cite{truncated_energy}) when this approach is taken. Avoiding this problem requires a segmented reconstruction that can measure the variations of energy loss along the track.

\subsection{Spatial Separation}
\label{sec:spatialdedx}

One approach to track segmentation is to assign each photomultiplier to one of several segments and then use the template-muon method (Eq. \ref{eq:poissonllh}) in each segment \cite{Mitsui92, truncated_energy}. This results in several averages taken over sections of the track, allowing the identification of outliers that distort the global average. Typically, these segments are $\sim 100\,\,\unit{m}$ in length, similar to the inter-string spacing in the IceCube array. When reduced to differentially small segments, each contains only one PMT; as they reach the size of the detector the result converges to that from the single-template method. This approach is described in more detail in \cite{truncated_energy}. An overview is given here.

\begin{figure}
\includegraphics[width=\linewidth]{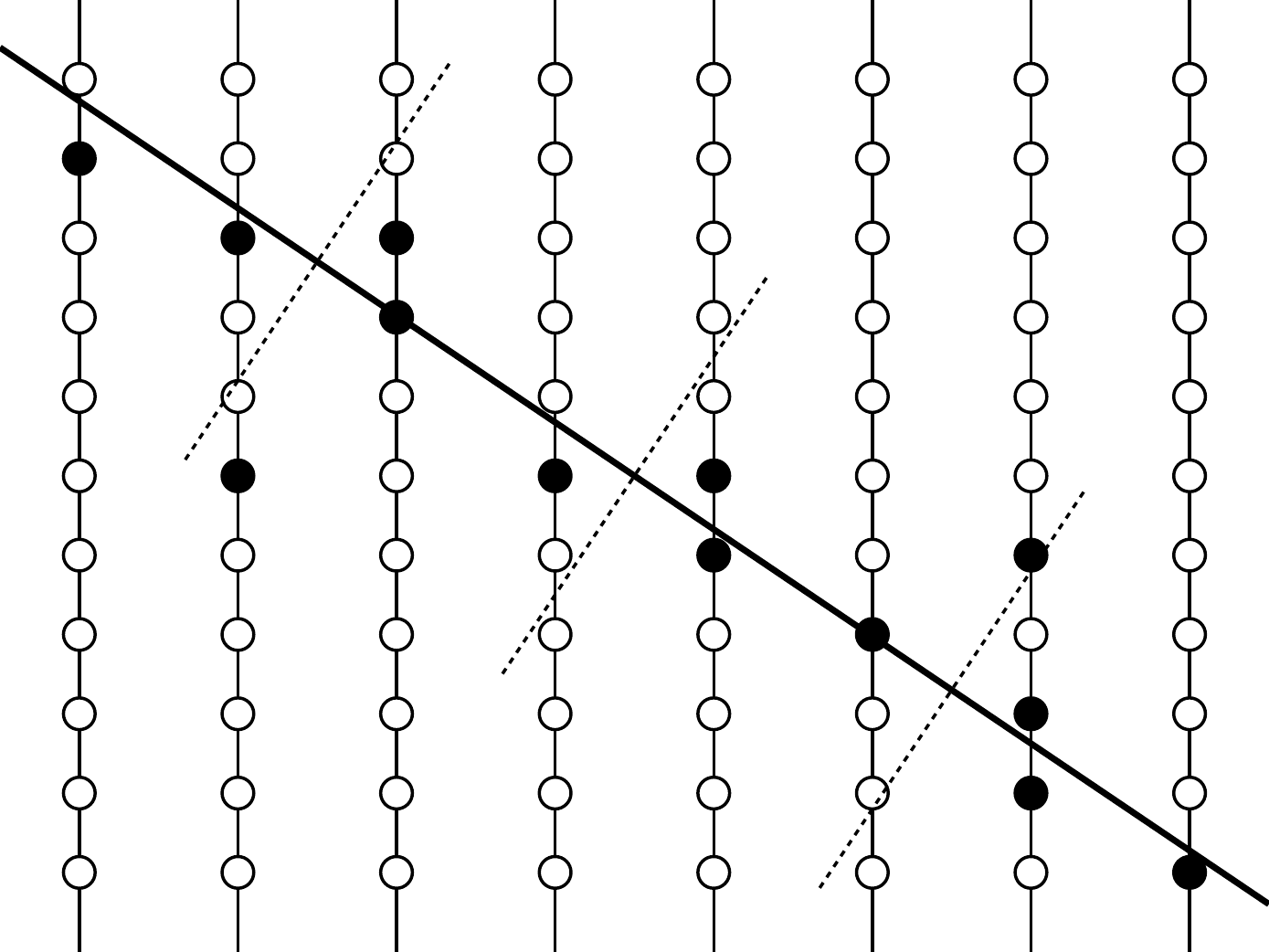}
\caption{Diagram of spatial binning of energy loss along a track, used in spatial-separation-based energy deposition reconstruction methods. Within each bin, separated by dashed lines, $\langle dE/dx \rangle$ is determined using a single template muon as though the photomultipliers in each bin constituted an independent detector.}
\label{fig:spatialbinning}
\end{figure}

When using spatial separation of the light deposition, the track is divided into bins bordered by planes perpendicular to the track, effectively treating each bin as a separate detector (Fig.~\ref{fig:spatialbinning}). The light from each PMT in the bin is treated in isolation from the remainder of the detector, and the energy of the track within the bin is determined using a single template muon following the same approach as for an unsegmented muon reconstruction.

Many factors influence the choice of bin size, with the objective to create bins as independent from each other as possible. IceCube uses a sparse grid of photomultipliers with sizable distances between PMTs, and each typically observes light from only a limited part of the muon track. The length of the optimal segmentation depends on the optical properties of the medium and expected correlation scale of the observed light. This is related to the expected energy of the muons: high-energy muons create more light that travels further through the ice, increasing the scale of correlations between segments. The typical absorption length for light in the IceCube array, averaged over the different optical properties at different depths in the ice sheet \cite{spice,aha}, is $\sim 125\,\,\unit{m}$. This is approximately the detector's inter-string spacing and is used as a typical value for the scale of track segmentation in spatial-separation-based reconstruction methods.

\subsection{Unfolding}
\label{sec:millipede}

Variations of muon energy loss can happen at scales much smaller than the usual segmentation scale used in spatial separation. Due to the lack of physical segmentation in the detector, light from single bright stochastic losses can travel distances longer than the size of the segments used and be detected simultaneously with photons from nearer parts of the muon track. Every PMT readout is then a combination of light from everywhere along the track emitted at many places within the detector.

The fact that this combination is linear, and that the individual stochastic losses take the form of electromagnetic showers (Sec.~\ref{sec:cascades}), makes unfolding these contributions tractable. We have already considered energy reconstruction in the presence of multiple overlaid light sources in Sec.~\ref{sec:energyreco} when discussing the inclusion of PMT noise and can generalize Eq.~\ref{eq:energyrecownoise} to allow the additional sources to be showers of variable energy by replacing the substitution

\begin{equation}
\lambda \rightarrow E \Lambda + \rho,
\end{equation}
with another where the expected photon count ($\lambda$) is the sum of photons from multiple sources ($\lambda_i$) and the noise rate ($\rho$):

\begin{equation}
\lambda \rightarrow \sum_{\textrm{sources}\,\, i} E_i \Lambda_i + \rho.
\end{equation}
Here each $E_i$ is the energy deposition by a particular sub-source $i$ and $\Lambda_i$ is the expected light yield in a particular photomultiplier and time bin from light source $i$. Our likelihood \eqref{eq:poissonllh} can then be rewritten in terms of vector operations:

\begin{equation}
\label{eq:millipedellh}
\begin{array}{rl}
\ln \mathcal{L}	& = k \ln \left (E_i \Lambda^i + \rho \right ) - E_i \Lambda^i - \rho - \ln \left (k! \right ) \\
& = k \ln \left ( \vec E \cdot \vec \Lambda + \rho \right ) - \vec E \cdot \vec \Lambda - \rho - \ln \left (k! \right ), \\
\end{array}
\end{equation}
and summing over time bins $j$:
\begin{equation}
\begin{array}{rll}
\sum_j \ln \mathcal{L} & = & \sum_j k_j \ln \left ( \vec E \cdot \vec \Lambda_j + \rho_j \right ) \\
 & & - \sum_j \left ( \vec E \cdot \vec \Lambda_j - \rho_j \right ) - \sum_j \left ( \ln k_j! \right ). \\
\end{array}
\end{equation}
Like Eq.~\ref{eq:energyrecownoise}, this has no analytic maximum for $\vec E$, but can be solved to first order (the approximately Gaussian error regime applicable at high energies):

\begin{equation}
k_j = \vec E \cdot \vec \Lambda_j + \rho_j.
\end{equation}

Introducing the matrix $\mathbf{\Lambda}$ for the predicted light yield at every point in the detector from every source position at some reference energy, this can be rewritten in terms of a matrix multiplication:

\begin{equation}
\vec k - \vec \rho = \mathbf{\Lambda} \cdot \vec E.
\label{eq:matrixmillipede}
\end{equation}

\begin{figure}
\begin{centering}
\includegraphics[width=\linewidth]{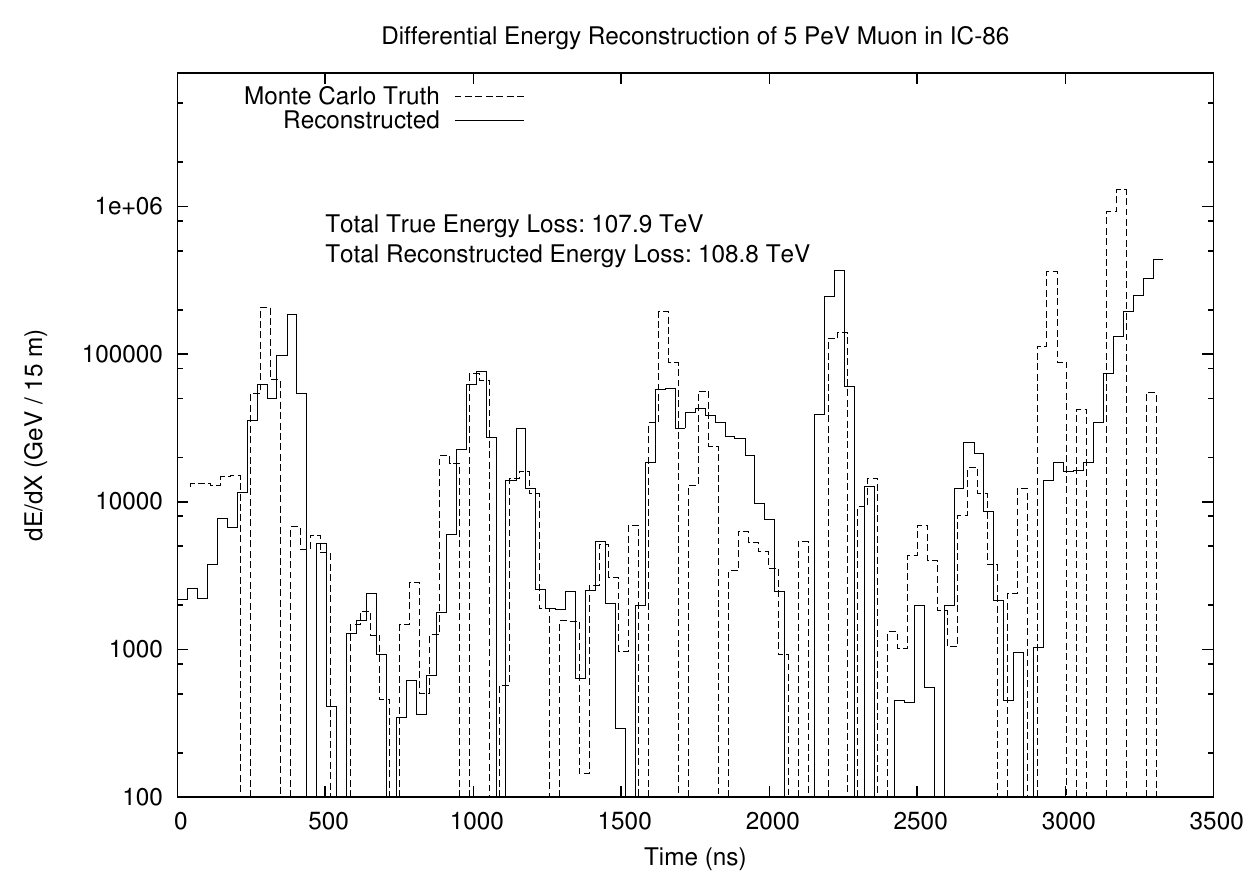}
\caption[Millipede $dE/dx$ reconstruction]{Reconstruction of the energy deposition of a simulated 5~PeV single muon using unfolding with a 15 meter cascade spacing. For this event, the total reconstructed energy loss within the detector differs from the true value by less than 1\%.}
\label{fig:millipedededx}
\end{centering}
\end{figure}

\begin{figure}
\begin{centering}
\begin{subfigure}{\linewidth}
\includegraphics[width=\linewidth]{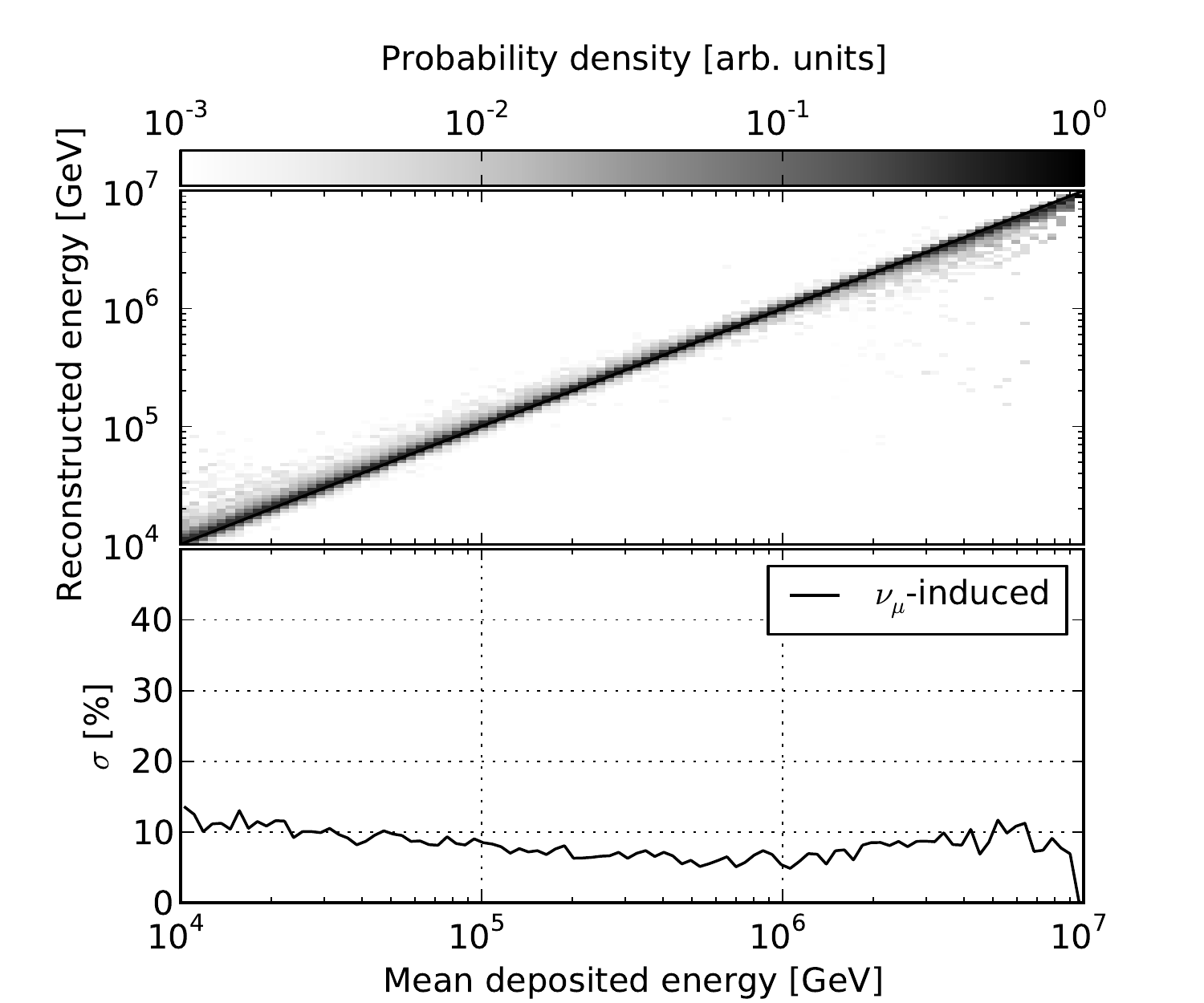}
\caption{Reconstructed total energy deposition as a function of true total energy deposition.}
\label{fig:millipede_deposition:2d}
\end{subfigure}
\begin{subfigure}{\linewidth}
\includegraphics[width=\linewidth]{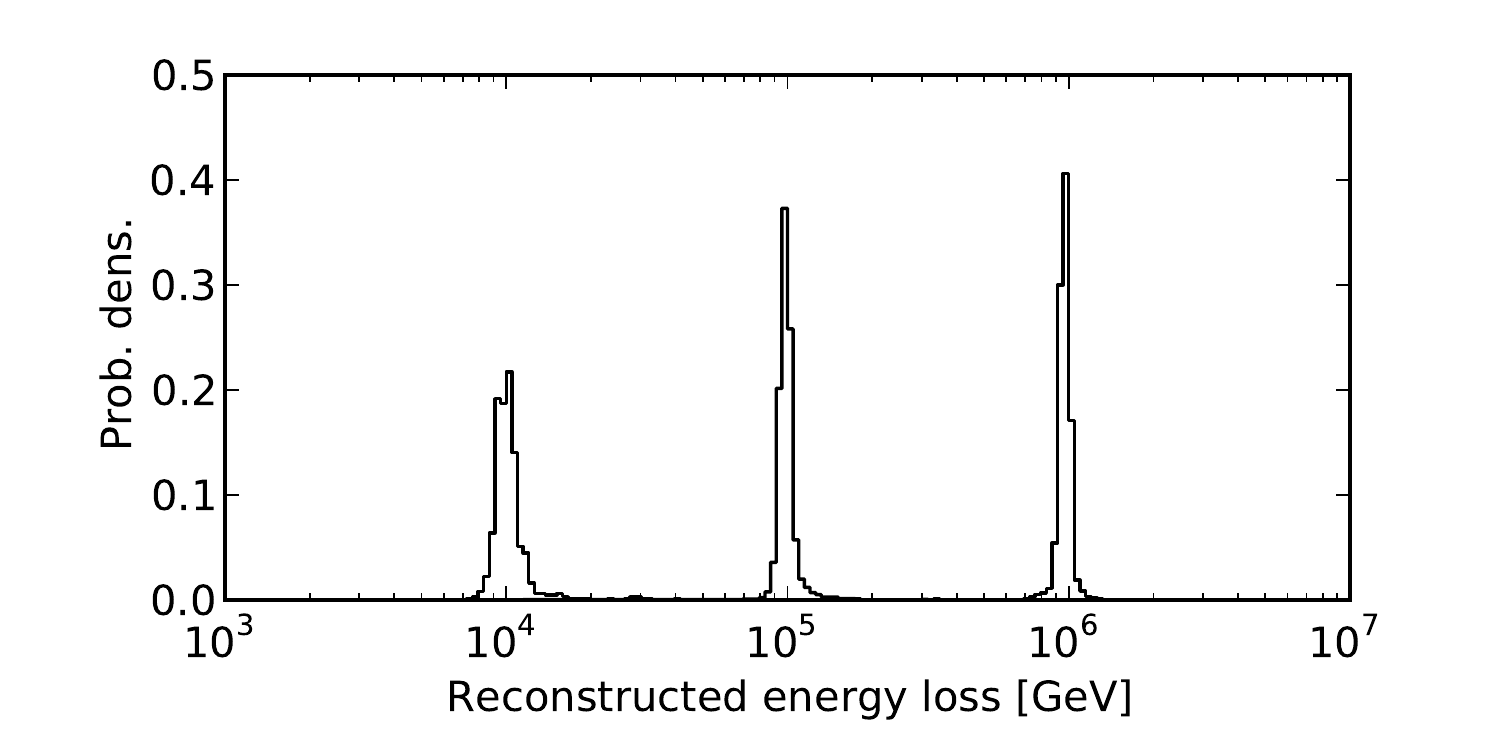}
\caption{Slices of the reconstructed total energy distribution shown in (\subref{fig:millipede_deposition:2d}) at fixed true total energy depositions of $10^4$, $10^5$, and $10^6$ GeV.}
\label{fig:millipede_deposition:slices}
\end{subfigure}
\caption[Energy resolution for $\nu_{\mu}$]{
Reconstruction, using unfolding with 2.5 meter cascade spacing, of the total energy deposition of simulated $\nu_{\mu}$ interactions with known directions and vertices that are inside the instrumented volume. Observation of such in-detector starting muon events allows positive identification as a charged-current $\nu_\mu$ interaction where all energy from the neutrino is deposited in the detector. High precision reconstruction of charged particle energies in such events then allows neutrino energy reconstruction limited only by instrumental resolution.
}
\label{fig:millipede_deposition}
\end{centering}
\end{figure}

Equation~\ref{eq:matrixmillipede} can be inverted by standard linear algebraic techniques to find the best-fit $\vec E$. We want, however, to incorporate additional physical constraints. In particular, negative energies are unphysical and should not be present in the solution even though a matrix inversion may often yield solutions with negative terms. The solution is to use a non-negative least squares algorithm \cite{Lawson:1974} to achieve a high-quality fit to the data (Fig.~\ref{fig:millipedededx}) by use of a linear deposition hypothesis with possible light sources every few meters along the track. This first-order linear solution can be further refined to exactly maximize Eq.~\ref{eq:millipedellh} by the use of algorithms used in positron emission tomography reconstructions. Here we use the Non-Monotonic Maximum Likelihood (NMML) algorithm \cite{nmml}, achieving resolution on total deposited energy along muon tracks of $\sim 10-15\%$ (Fig. \ref{fig:millipede_deposition}), comparable to that achieved for deposited energies with cascade events (Fig.~\ref{fig:monopod_performance}).

Additional physical constraints can be included by the use of regularization terms in the likelihood \eqref{eq:millipedellh}. Although most uses for regularization (preventing ringing, in particular) are eliminated by the non-negativity constraint, such terms can still be useful as additional weak penalties. We use Tikonoff regularization to accomplish two goals.  Adding a term proportional to the norm of the first derivatives of the $dE/dx$ vector (a first-order penalty) can be used to restrict fluctuations in muon energy loss. More commonly we apply an extremely weak ridge penalty on $||dE/dx||$ to break inherent degeneracies in the solution to Eq.~\ref{eq:millipedellh} between nearby low-energy events and distant high-energy events when only small numbers of photons on the detector boundary are observed. In such cases, the weak penalty causes the fit to prefer the nearby, low-energy solution.

\section{Interpretation of Segmented Energy Losses}

\begin{figure}
\begin{centering}
\begin{subfigure}{\linewidth}
\includegraphics[width=\linewidth]{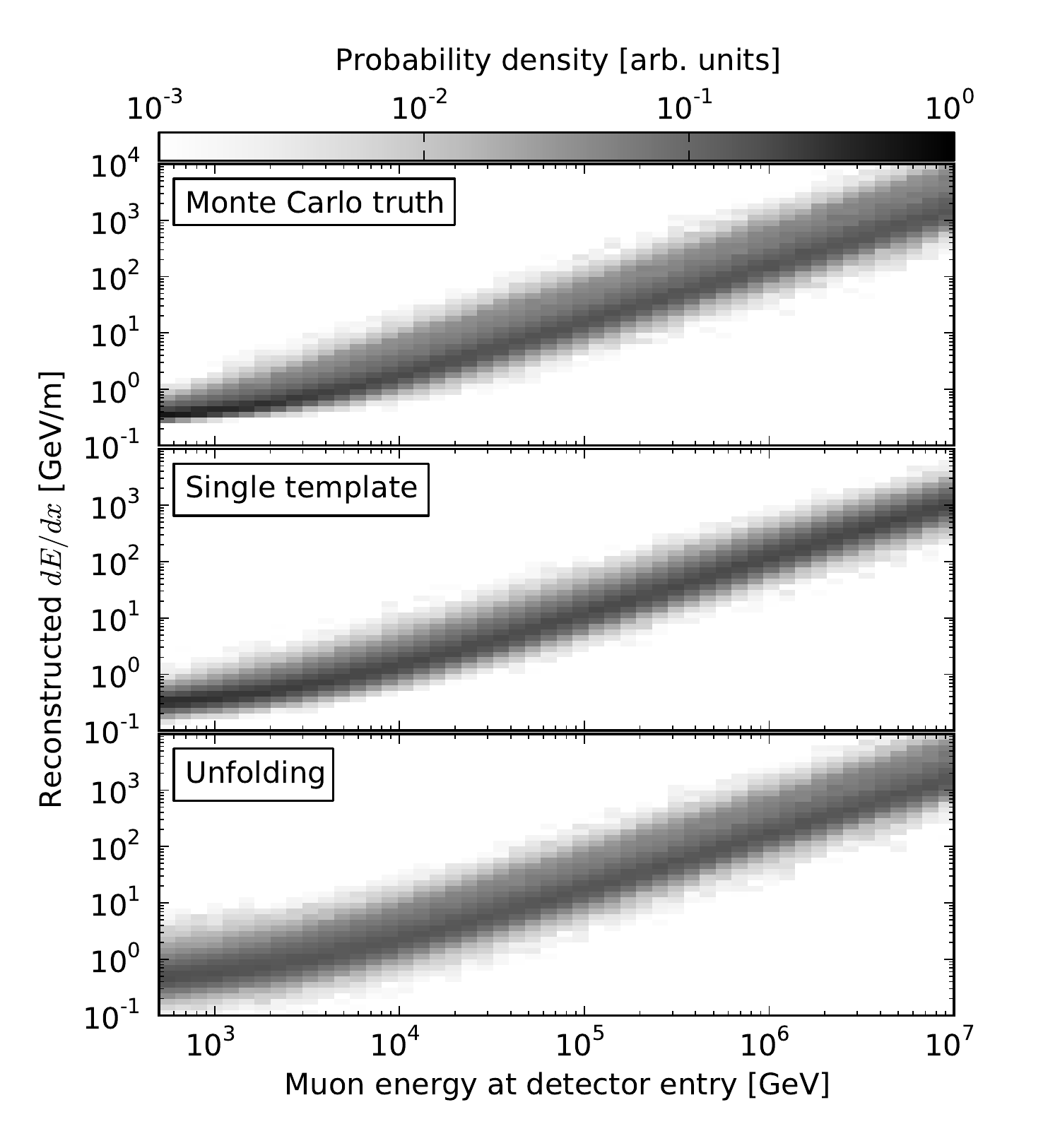}
\caption{Mean energy loss rate determined from true Monte Carlo information (top panel), the single-muon template method (middle) panel, and multi-source unfolding (bottom panel).
}
\label{fig:dEdx_v_E_zoo:2d}
\end{subfigure}
\begin{subfigure}{\linewidth}
\includegraphics[width=\linewidth]{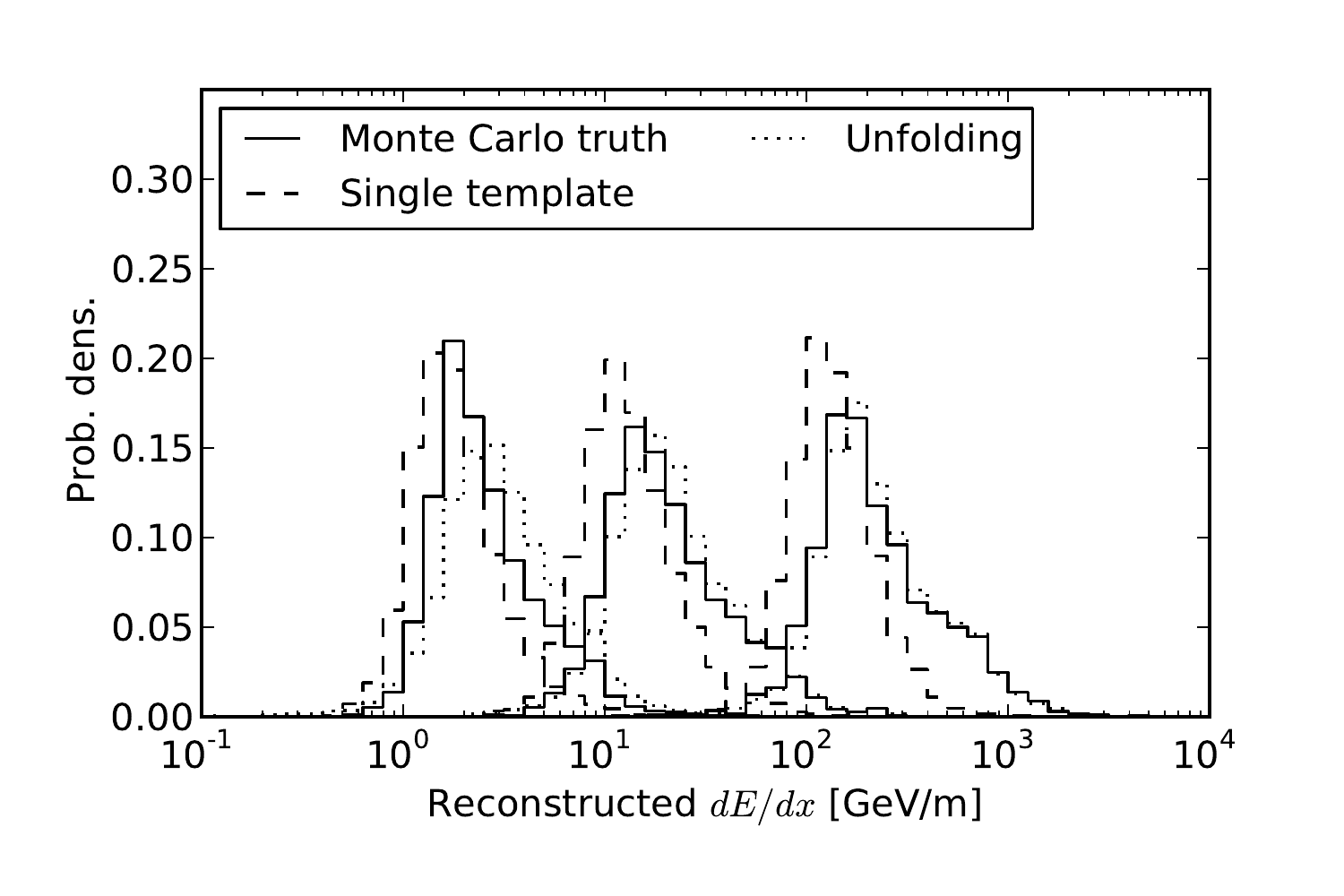}
\caption{Slices of the reconstructed energy loss rate distributions shown in (\subref{fig:dEdx_v_E_zoo:2d}) at fixed muon energies of $10^4$, $10^5$, and $10^6$ GeV.}
\label{fig:dEdx_v_E_zoo:slices}
\end{subfigure}
\caption{Mean energy loss rate of through-going muons as a function of the muon energy at the point where it enters the detector volume.
While the unfolding method reproduces well the fluctuations in the true energy loss rate above a few TeV (the similarity in $\langle dE/dx \rangle$ between ``Monte Carlo truth'' and ``Unfolding'') , these fluctuations limit the usefulness of the mean loss rate as a proxy for the energy of through-going muons.
}
\label{fig:dEdx_v_E_zoo}
\end{centering}
\end{figure}

The mean energy loss rate of a muon ($\langle dE/dx \rangle$) is roughly proportional to its energy above $\sim 1\,\,\unit{TeV}$, and its accurate measurement is thus the focus of most existing IceCube muon energy reconstructions. However, IceCube can only observe a fraction of the muon track at these energies and therefore statistically robust methods for the estimation of $\langle dE/dx \rangle$ must be used. A simple average (top panel of Fig.~\ref{fig:dEdx_v_E_zoo}) provides poor muon energy resolution with large non-Gaussian tails due to statistical bias from large stochastic losses in the detector. The segmented energy loss rates computed in the previous section, however, can be used to develop more robust estimators. The most common approach is to use the truncated mean instead of a straight average to reduce the effects of outliers; other techniques that use information about the likelihood of large losses may further improve resolution.

\subsection{Measuring the resolution of an energy proxy}

% Jakob, inspired by Gary

\begin{figure}
\begin{centering}
\includegraphics[width=\linewidth]{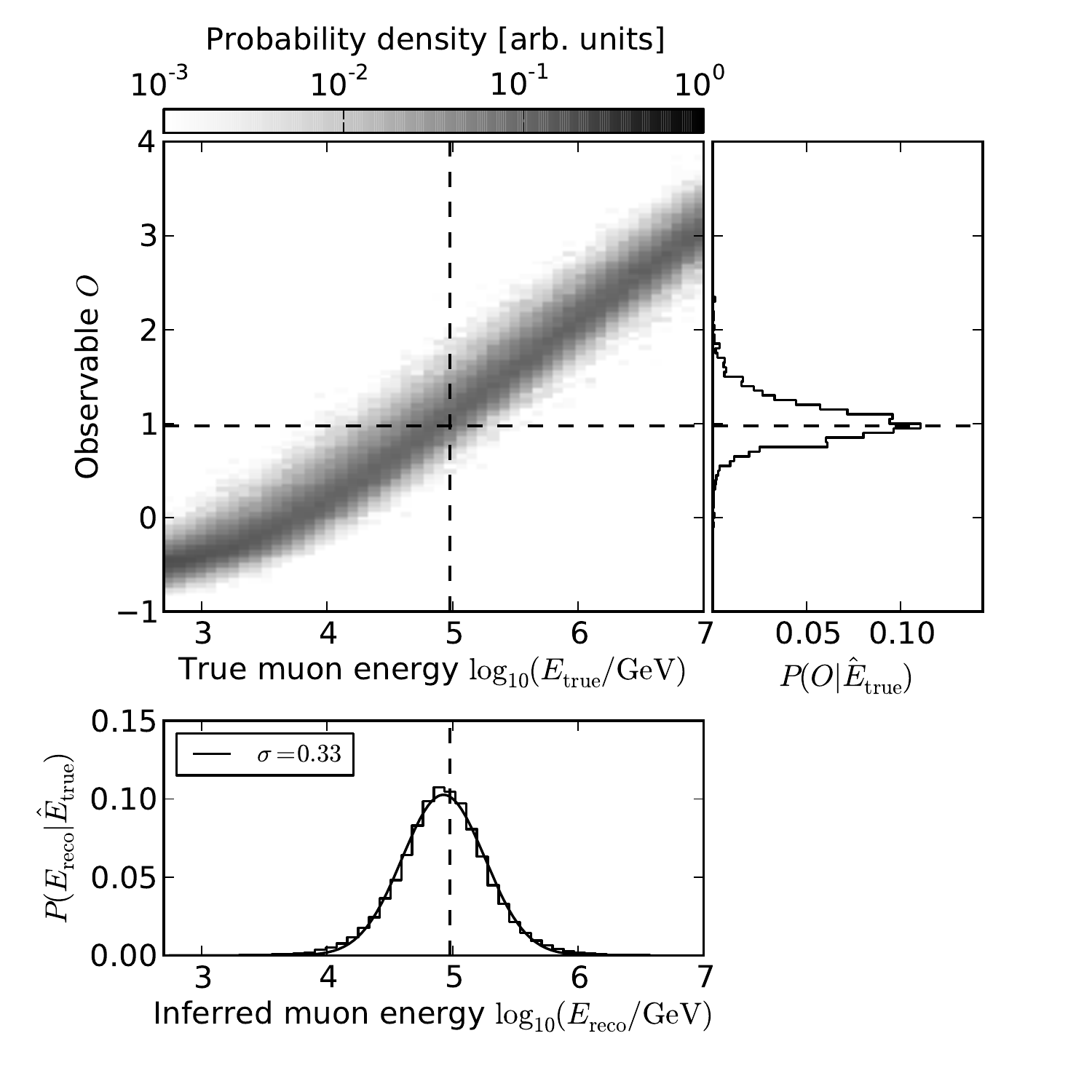}
\caption{A construction for determining the energy resolution of a proxy observable. Muons of a given energy $\hat E_{\rm true}$ (vertical dashed line at $\log_{10}(E_{\rm true}/\unit{GeV}) = 5$) will produce a distribution $P(O|\hat E_{\rm true})$ of the proxy observable $O$ (right panel). Each $O$, in turn, has an associated distribution of true energies $P(E_{\rm true}|O)$. The distribution of best-fit energies $E_{\rm reco}$ for muons with a given $\hat E_{\rm true}$, shown as a solid line in the bottom panel, is given by $\int_O P(E_{\rm true}|O) P(O|\hat E_{\rm true})dO$. The width of this distribution is a measure of the energy resolution of the proxy observable (in this case, $\log_{10}(dE/dx)$ from the ``single template'' method) for muons of the given energy.}
\label{fig:dEdx_resolution_construction}
\end{centering}
\end{figure}

All the observables discussed here (total energy loss, mean energy loss rate, truncated mean energy loss rate) are related to the energy of the underlying muon but are not themselves energies. In order to be able to discuss the resolving power of these observables, we require a way to measure the resolution of proxy observables with disparate ranges and units. We can construct such a measure for each proxy observable by simulating events and building up a joint distribution of muon energy and the proxy observable like the one shown in Fig.~\ref{fig:dEdx_resolution_construction}. The relationship between the muon energy and the observable is not one-to-one. The proxy observable can take on a wide range of values for muons of the same energy, and each value of the observable can in turn arise from a range of muon energies. This ambiguity can be captured by constructing a confidence interval as shown in Fig.~\ref{fig:dEdx_resolution_construction}. In essence this confidence interval measures the vertical width of the observable distribution in proportion to the overall slope of the joint distribution. In the minimum-ionizing regime, where the energy loss rate becomes nearly independent of energy, the distribution must be quite narrow in order to separate muons of different energies. At high energies, however, it can become wider while maintaining an equivalent resolving power.

\subsection{Truncated Mean}

% Marius

By discarding energy losses from segments of the muon track with the highest loss rates, it is possible to obtain a more robust measurement of the typical $\langle dE/dx \rangle$ of the track (the truncated mean approach \cite{1994NIMPA.343..463A,truncated_energy}). Rejecting the bins with the largest energy losses removes outliers from the average and the variance in the calculation of $\langle dE/dx \rangle$ and the muon energy is therefore reduced. This approach can be applied to the results of segmented $dE/dx$ reconstructions using both spatial separation (Sec.~\ref{sec:spatialdedx}, \cite{truncated_energy}) and unfolding (Sec.~\ref{sec:millipede}).

\begin{figure}
\begin{centering}
\includegraphics[width=\linewidth]{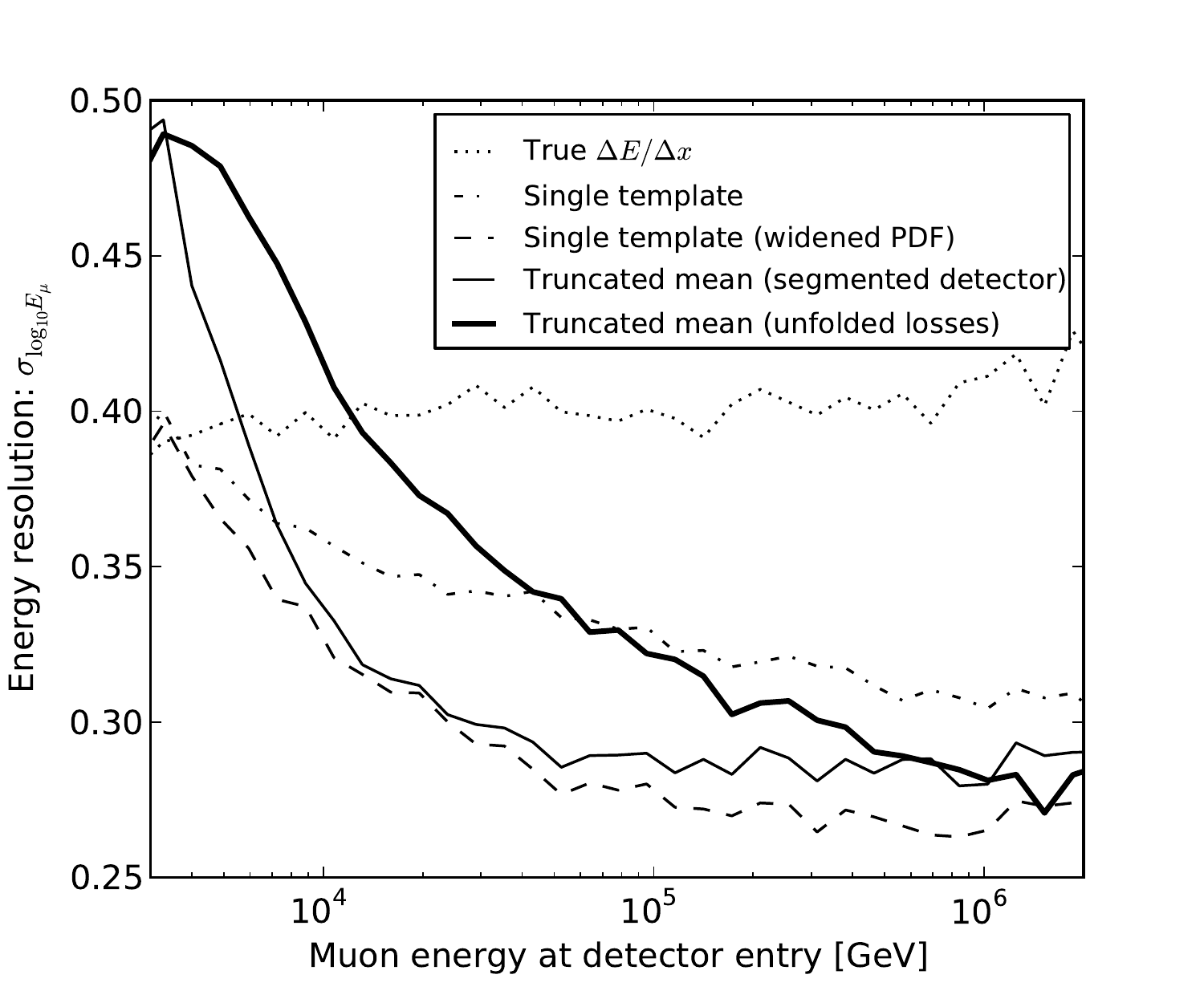}
\caption[]{Comparison of the energy resolution of various methods on simulated through-going muons using the measure illustrated in Fig.~\ref{fig:dEdx_resolution_construction}. The true mean energy loss rate is shown for purposes of comparison only and provides a poor energy proxy no matter how precisely it can be reconstructed.}
\label{fig:dEdx_resolution}
\end{centering}
\end{figure}

\begin{figure}
\begin{centering}
\includegraphics[width=\linewidth]{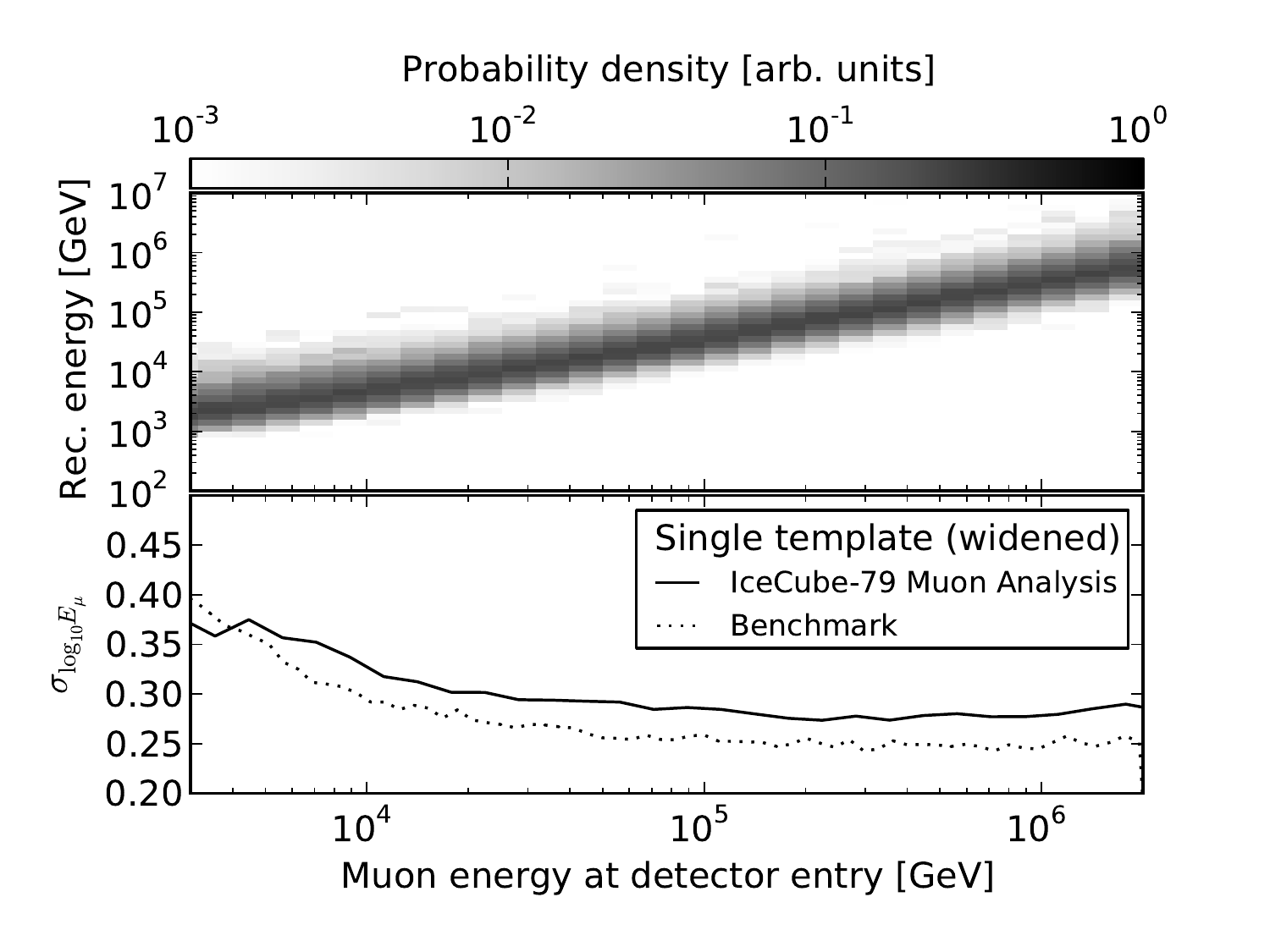}
\caption[]{Energy resolution for a sample IceCube muon analysis at final level. The slight reduction in resolution at high energies compared to the ``benchmark'' level (from Fig.~\ref{fig:dEdx_resolution}) arises from convolution of the intrinsic energy resolution (``benchmark'') with the finite resolution of the angular and positional reconstruction used in this analysis.}
\label{fig:cweaver_diffuse}
\end{centering}
\end{figure}

This reconstruction technique provides substantially better estimates of the muon energy than deposition-only or single-muon-template estimates using either segmented measurement of $dE/dx$ (Fig. \ref{fig:dEdx_resolution}). Similar performance can also be achieved by widening the upper end of the photon counting probability densities (Sec.~\ref{sec:analytic_approximation}, here with $w = 10$) in the single-template method to achieve an effective truncated mean (``single template (widened PDF)'' in Fig. \ref{fig:dEdx_resolution}). Although these resolutions are computed in an idealized case, in which the muon position and direction are known exactly, typical resolutions at the analysis level, given standard muon geometry reconstructions \cite{2004NIMPA.524..169A}, exhibit similar behavior (Fig.~\ref{fig:cweaver_diffuse}).

\subsection{Future Topics}

Analysis of energy loss topologies is currently a subject of ongoing work and will be discussed in detail in future publications. Two possible applications of this information are discussed below.

% Gary

When treated in detail, the rate and amplitude of high-energy stochastic losses may provide more detail about muon energy than a $\langle dE/dx \rangle$ measurement \cite{Mitsui92}. The most interesting case is muons emitted from in-detector $\nu_\mu$ interactions where all charged particles are observed and unambiguous flavor identification is possible. Accurate determination of the energy of the outgoing muon, taking into account the event topology, may allow neutrino energy resolution for these events to approach the deposited energy resolution (Fig. \ref{fig:millipede_deposition}).

% Marcel, Patrick?

\begin{figure}
\begin{centering}
\begin{subfigure}{\linewidth}
	\includegraphics[width=\linewidth]{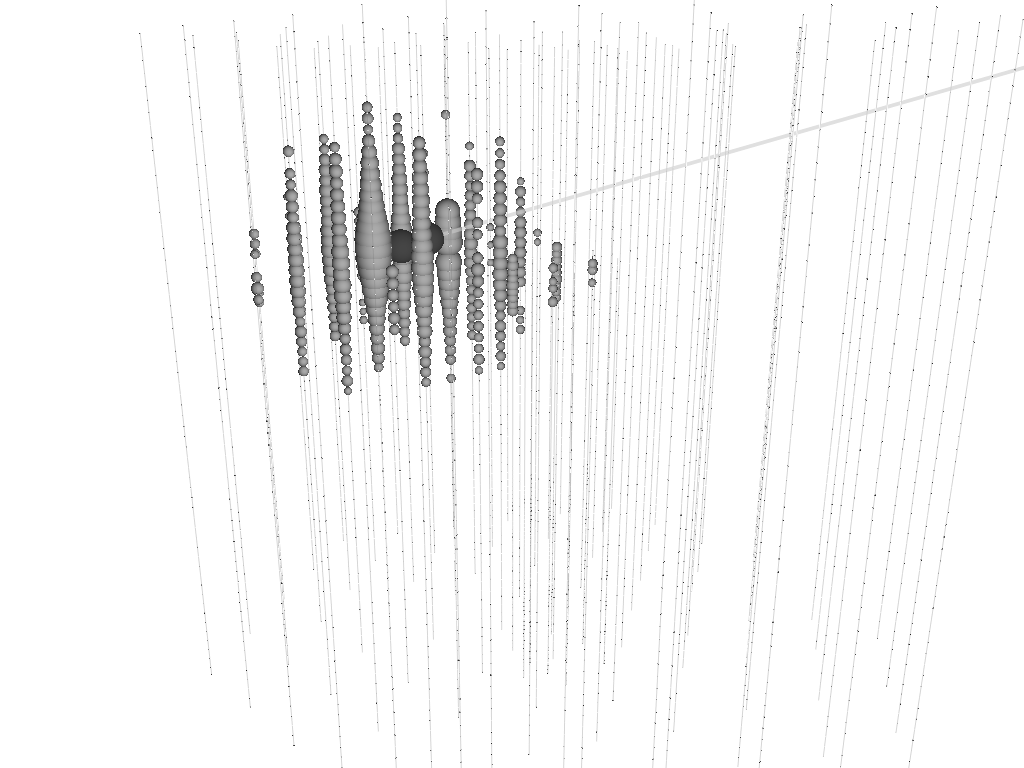}
	\caption{A simulated charged-current $\nu_\tau$ interaction creating a hadronic cascade and a $\tau$ which decays after 50\,m to a second cascade. The black spheres indicate the energy depositions of the two cascades. On a macroscopic level, this event looks very cascade-like.}
	\label{fig:double_bang_steamshovel}
\end{subfigure}
\begin{subfigure}{\linewidth}
\includegraphics[width=\linewidth]{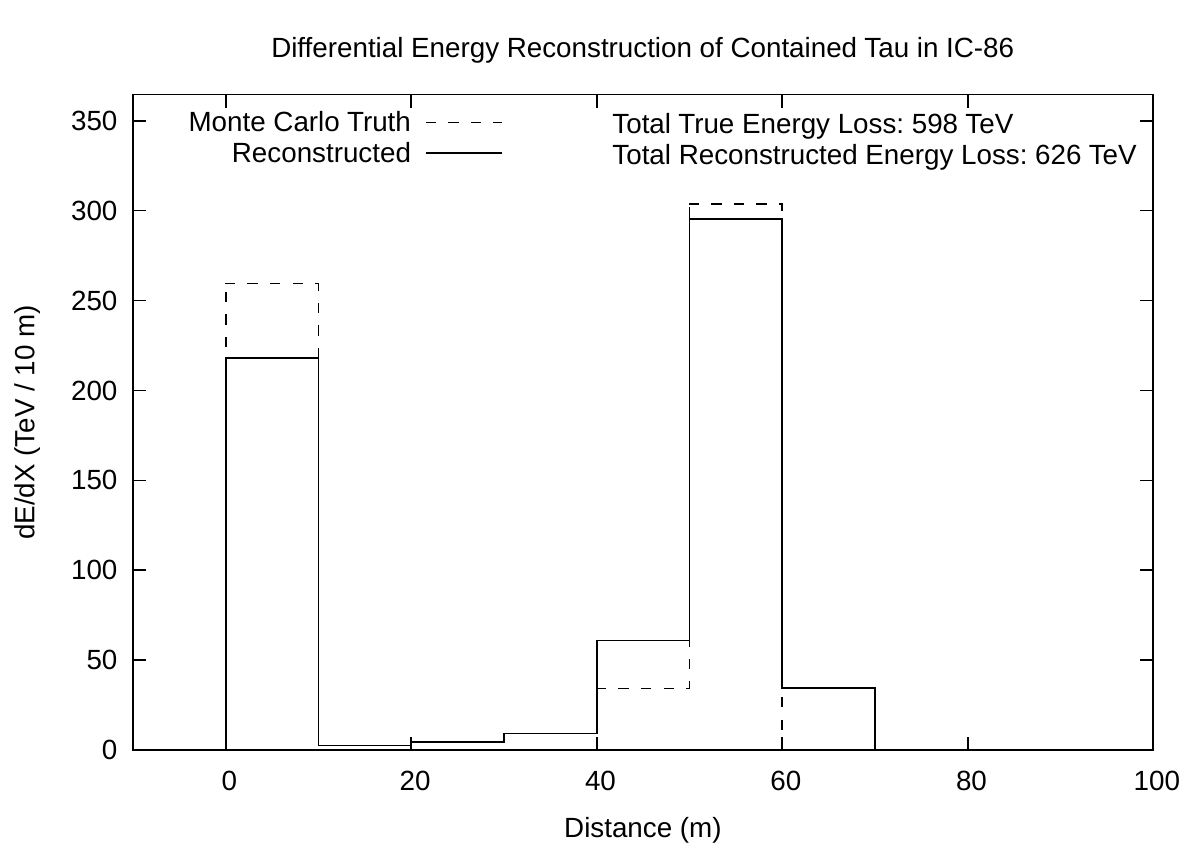}
\caption[$dE/dx$ reconstruction of CC $\nu_{\tau}$]{Differential energy reconstruction showing two separated peaks, a distinct signature of $\nu_{\tau}$ interactions. Detailed $dE/dx$ reconstruction may allow identification of these interactions with decay lengths smaller than the scale of the detector instrumentation.}
\label{fig:millipede_double_bang}
\end{subfigure}
\caption[Example of $\nu_\tau$ interaction]{Example of a simulated charged-current $\nu_\tau$ interaction with subsequent decay producing a second cascade (also known as ``Double Bang''; see Table~\ref{table:signatures}).}
\label{fig:double_bang}
\end{centering}
\end{figure}

Detailed energy loss topologies are also useful beyond muon energy measurements for identification of interactions and determination of neutrino flavor. Electron neutrino charged-current interactions and all neutral-current interactions have nearly point-like energy deposition (Sec.~\ref{sec:cascades}), while muons have extended tracks, potentially beginning with a bright hadronic cascade at the neutrino vertex. Charged-current $\nu_\tau$ interactions can have a variety of signatures, with tracks from taus (length $\sim 50\,\,\unit{m} \left ( \frac{E}{1\,\,\unit{PeV}} \right )$) and potential decay muons as well as hadronic cascades associated with the neutrino vertex and tau decay. High-resolution segmented $dE/dx$ reconstruction, as developed here, may allow positive identification of these type of interactions even when the $\tau$ track has a length shorter than the detector instrumentation spacing, which results in an otherwise cascade-like light pattern in the detector (Fig.~\ref{fig:double_bang}).

\section{Conclusion}

Using the techniques described here, IceCube achieves average deposited energy resolution in all channels of $\sim 15\%$ (depending on the event selection used in individual analyses, better resolutions are possible). This is limited primarily by systematic uncertainties above a few TeV. Some of these may be reduced in the future, in particular those due to modeling of light propagation and shower development. Others, such as the inherent variance of hadronic shower light yield, are fundamental limitations to IceCube's performance.
Deposited energy resolution is highest for near-pointlike largely electromagnetic particle showers, such as those produced in $\nu_e$ interactions, for which resolution is limited by shower fluctuations over most of IceCube's energy range.

For extended objects, such as muon tracks, precision reconstruction and localization of energy loss topology are possible.
Work is on-going to use this information fully for high-quality muon energy reconstruction, in particular in the case of muons produced in $\nu_\mu$ charged-current interactions, where neutrino flavor can be measured directly and all charged particles are visible.
As this work on using the detailed energy loss patterns of muons to measure their energies continues, a variety of alternative observables can be used, optimized for different use cases.
Searches for muon neutrinos interacting within the detector, where the visible energy is dominated by the contained hadronic cascade at the vertex, typically use integrated deposited energy as an observable, as in \cite{hese_paper} and Fig.~\ref{fig:millipede_deposition}.
For searches focused on higher energy uncontained events, the cascade at the vertex is unobservable and muon energy is a more robust indicator of neutrino energy than the deposited energy in the detector.
Such analyses \cite{ic59_muons} typically use one of the stochastic-loss-filtering muon energy estimators shown in Fig.~\ref{fig:dEdx_resolution}.
A wealth of additional information is provided by these topological reconstructions and the likelihood model described here: in addition to reconstruction of muon energies, topologies can also be used to study muon energy loss processes at very high energies and for particle identification.

\section*{Acknowledgments}
We acknowledge support from the following agencies:
U.S. National Science Foundation-Office of Polar Programs, U.S. National Science Foundation-Physics Division, University of Wisconsin Alumni Research Foundation, the Grid Laboratory Of Wisconsin (GLOW) grid infrastructure at the University of Wisconsin - Madison, the Open Science Grid (OSG) grid infrastructure; U.S. Department of Energy, and National Energy Research Scientific Computing Center, the Louisiana Optical Network Initiative (LONI) grid computing resources; Natural Sciences and Engineering Research Council of Canada, WestGrid and Compute/Calcul Canada; Swedish Research Council, Swedish Polar Research Secretariat, Swedish National Infrastructure for Computing (SNIC), and Knut and Alice Wallenberg Foundation, Sweden; German Ministry for Education and Research (BMBF), Deutsche Forschungsgemeinschaft (DFG), Helmholtz Alliance for Astroparticle Physics (HAP), Research Department of Plasmas with Complex Interactions (Bochum), Germany; Fund for Scientific Research (FNRS-FWO), FWO Odysseus programme, Flanders Institute to encourage scientific and technological research in industry (IWT), Belgian Federal Science Policy Office (Belspo); University of Oxford, United Kingdom; Marsden Fund, New Zealand; Australian Research Council; Japan Society for Promotion of Science (JSPS); the Swiss National Science Foundation (SNSF), Switzerland; National Research Foundation of Korea (NRF); Danish National Research Foundation, Denmark (DNRF). 
N.W. was supported by the NSF GRFP.

\bibliographystyle{elsarticle-num}

\end{document}